\documentclass[12pt, preprint]{revtex4-2}

\usepackage{slashed}
\usepackage{bbm} 
\usepackage{bm} 
\usepackage{mathrsfs}
\usepackage{amsfonts}
\usepackage{amssymb}
\usepackage{amsmath}
\usepackage{amsthm}
\usepackage{mathtools}

\usepackage{color}
\definecolor{myblue}{rgb}{.8, .8, 1}

\usepackage{empheq}
\newlength\mytemplen
\newsavebox\mytempbox

\makeatletter
\newcommand\mybluebox{%
    \@ifnextchar[
       {\@mybluebox}%
       {\@mybluebox[0pt]}}

\def\@mybluebox[#1]{%
    \@ifnextchar[
       {\@@mybluebox[#1]}%
       {\@@mybluebox[#1][5pt]}}

\def\@@mybluebox[#1][#2]#3{
    \sbox\mytempbox{#3}%
    \mytemplen\ht\mytempbox
    \advance\mytemplen #1\relax
    \ht\mytempbox\mytemplen
    \mytemplen\dp\mytempbox
    \advance\mytemplen #2\relax
    \dp\mytempbox\mytemplen
    \colorbox{myblue}{\hspace{1em}\usebox{\mytempbox}\hspace{1em}}}

\makeatother

\numberwithin{equation}{section}

\theoremstyle{definition}
\newtheorem{definition}{Definition}[section]

\theoremstyle{remark}

\usepackage{braket}

\usepackage{mathtools}

\usepackage{physics}

\usepackage{tensor}

\usepackage{hyperref}
\hypersetup{
     colorlinks=true, 
     linktoc=all,     
     linkcolor=black,  
 }

\renewcommand{\bra}[1]{\langle #1\vert}
\renewcommand{\ket}[1]{\vert #1 \rangle}
\renewcommand{\braket}[2]{\langle #1 \vert #2 \rangle}

\DeclareMathAlphabet{\mathbb}{U}{msb}{m}{n}

\newcommand{\rn}{\mathbb{R}}

\newcommand{\cn}{\mathbb{C}}

\newcommand{\id}{\mathbbm{1}}

\newcommand{\df}{\coloneqq}

\newcommand{\glc}{\mathfrak{gl}(2,\cn)}

\newcommand{\slc}{\mathfrak{sl}(2,\cn)}

\newcommand{\so}{\mathfrak{so}(1,3)}

\newcommand{\soc}{\mathfrak{so}(1,3)_{\cn}}

\newcommand{\GL}{\text{GL}(2,\cn)}

\newcommand{\SLC}{\text{SL}(2,\cn)}

\newcommand{\SO}{\text{SO}(1,3)}

\newcommand{\SOP}{\text{SO}(1,3)^\uparrow}

\newcommand{\p}[1]{\sigma_{#1}}

\newcommand{\J}[1]{J_{#1}}

\newcommand{\re}[3]{\rho_{#3}(#1_{#2})}

\newcommand{\hre}[3]{\rho_{#3}(#1_{#2})}

\newcommand{\hr}[1]{\rho_{#1}}

\newcommand{\hf}{\frac{1}{2}}

\newcommand{\dr}[2]{\kappa_{#2} (#1)}

\newcommand{\lr}[1]{D_{#1} (\Lambda)}

\newcommand{\plr}[1]{D^{\textsc{H}}_{#1} (\lt)}

\newcommand{\td}[1]{\tilde{#1}}

\newcommand{\al}{\alpha}

\newcommand{\bt}{\beta}

\newcommand{\gm}{\gamma}

\newcommand{\ka}{\kappa}

\newcommand{\ep}{\varepsilon}

\newcommand{\hil}{\mathcal{H}}

\newcommand{\lag}{\mathscr{L}}

\newcommand{\fs}{\mathcal{F}}

\newcommand{\cv}{\mathcal{V}}

\newcommand{\pd}{\braket{\cdot}{\cdot}}

\newcommand{\hp}{\hat{\pi} }

\newcommand{\heta
}{\hat{\eta}}

\newcommand{\hps}{\hat{\psi}}

\newcommand{\hpsd}{\hat{\psi}^{\dag}}

\newcommand{\hph}{\hat{\phi}}

\newcommand{\hphd}{\hat{\phi}^{\dag}}

\newcommand{\hu}{\hat{U}}

\newcommand{\hpm}{\hat{P}}

\newcommand{\hJ}{\hat{J}}

\newcommand{\hpr}{\hat{\mathcal{P}}}

\newcommand{\htr}{\hat{\mathcal{T}}}

\newcommand{\thps}[2]{\hat{\td{\psi}}^{#1} (#2)}

\newcommand{\thph}[3]{\hat{\td{\phi}}^{#1}_{#2} (#3)}

\newcommand{\hC}{\hat{\mathcal{C}}}

\newcommand{\ha}[2]{\hat{a}^{#1}(0,#2)}

\newcommand{\hda}[2]{\hat{a}^{\dag #1}(0,#2)}

\newcommand{\hc}[2]{\hat{c}^{#1}(0,#2)}

\newcommand{\hdc}[2]{\hat{c}^{\dag #1}(0,#2)}

\newcommand{\hA}{\hat{A}}

\newcommand{\hh}{\hat{H}}

\newcommand{\hlag}{\hat{\lag}}

\newcommand{\vx}{\vec{x}}

\newcommand{\vp}{\vec{p}}

\newcommand{\lt}{\Lambda}

\newcommand{\der}[3]{\partial_{#1}^{#2}{#3}}

\newcommand{\be}{\begin{equation}}

\newcommand{\ee}{\end{equation}}

\newcommand{\bi}{\begin{itemize}}

\newcommand{\ei}{\end{itemize}}

\newcommand{\comma}{`}

\begin{document}

\setlength{\abovedisplayskip}{12pt}
\setlength{\belowdisplayskip}{12pt}

\title{Poincar\'{e} symmetries and representations in \\ pseudo-Hermitian quantum field theory}
\author{Esra Sablevice}
\email{esra.sablevice@manchester.ac.uk}
\author{Peter Millington}
\email{peter.millington@manchester.ac.uk}
\affiliation{Department of Physics and Astronomy, University of Manchester,\\ Manchester M13 9PL, United Kingdom}
\date{3 April 2024}

\begin{abstract}
This paper explores quantum field theories with pseudo-Hermitian Hamiltonians, where $PT$-symmetric Hamiltonians serve as a special case. In specific regimes, these pseudo-Hermitian Hamiltonians have real eigenspectra, orthogonal eigenstates, and unitary time evolution. So far, most pseudo-Hermitian quantum field theories have been constructed using analytic continuation or by adding non-Hermitian terms to otherwise Hermitian Hamiltonians. However, in this paper, we take a different approach. We construct pseudo-Hermitian scalar and fermionic quantum field theories from first principles by extending the Poincaré algebra to include non-Hermitian generators. This allows us to develop consistent pseudo-Hermitian quantum field theories, with Lagrangian densities that transform appropriately under the proper Poincaré group. By doing so, we establish a more solid theoretical foundation for the emerging field of non-Hermitian quantum field theory.
\\
~
\\
\footnotesize{This is an author-prepared post-print of \href{https://doi.org/10.1103/PhysRevD.109.065012}{Phys.\ Rev.\ D {\bf 109} (2024) 065012}, published by the American Physical Society under the terms of the \href{https://creativecommons.org/licenses/by/4.0/}{CC BY 4.0} license (funded by SCOAP\textsuperscript{3}).}

\end{abstract}

\maketitle

\newpage

\tableofcontents

\newpage


\section{Introduction}

In ``standard'' quantum mechanics, the description of a physical system relies on two crucial elements:\ a Hilbert space $\hil$ of states and a Hamiltonian operator $\hh: \hil \mapsto \hil$ that determines the time evolution. The Hamiltonian and real-valued physical observables correspond to Hermitian operators. This guarantees that their expectation values are real and that their eigenstates are orthogonal. Moreover, the time evolution operator generated by a Hermitian Hamiltonian is unitary, ensuring that probability is conserved.

However, it is now well established that operators do not need to be Hermitian to produce real expectation values~\cite{Bender:1998ke, Bender:1998gh, Mostafazadeh:2001jk, Mostafazadeh:2001nr}. Instead, an operator, say $\hat{A}$, must satisfy a condition known as \emph{pseudo-Hermiticity}:\:\ $\hat{A}^{\dag}=\heta \hat{A}\heta^{-1}$~\cite{Mostafazadeh:2008pw} (where $\dag$ is the usual composition of complex conjugation and matrix transposition) for some Hermitian operator $\heta^{\dag}=\heta$. We define a new inner product $\braket{\cdot}{\cdot}_{\heta}\df\braket{\cdot}{\heta\cdot}$, which yields real expectation values of $\hA$. Unlike Hermitian operators, whose eigenvalues are always real, pseudo-Hermitian operators exhibit eigenvalues that are either real, come in complex-conjugate pairs, or so-called exceptional points, where eigenvalues merge and the operator becomes defective~\cite{Heiss_2004}. When the eigenvalues of the Hamiltonian are real, there exists an additional discrete symmetry of the Hamiltonian that ensures unitary time evolution~\cite{Bender:2002vv}, and the pseudo-Hermitian theory can be made Hermitian via a similarity transformation~\cite{Mostafazadeh:2002id}. However, the similarity transformation becomes singular at the exceptional points, and there is no Hermitian counterpart when the eigenenergies are complex~\cite{BenderBook, Bender:2005tb}.

While non-Hermitian quantum mechanics has been studied extensively and applied to various physical systems (for a review, see Ref.~\cite{El-Ganainy:2018ksn}) across optics \cite{Zyablovsky_2014, Wen:2018vjx, Ashida:2020dkc}, photonics~\cite{Ruschhaupt_2005, Klaiman:2008zz}, and condensed matter physics~\cite{Dembowski:2001zz, Heiss:2012dx}, the subject of non-Hermitian quantum field theory has only recently gained traction and is still a developing field. To date, much of the development in non-Hermitian quantum field theory has focused on simplified models of $PT$-symmetric quantum field theories. $PT$-symmetric quantum field theories~\cite{Bender:2005tb} are a subset of pseudo-Hermitian quantum field theories, in which the Hamiltonian is symmetric under the combined action of parity $P$ and time-reversal $T$. Examples include the following:\ $i\phi^3$ scalar quantum field theory~\cite{Bender:2004vn, Bender:2004sa, Bender:2012ea, Bender:2013qp}, the ``wrong-sign'' $-\phi^4$ theory~\cite{Shalaby:2006fh, Shalaby:2009xda, Ai:2022csx}, and generalizations~\cite{Bender:2018pbv}; non-Hermitian extensions of the Dirac Lagrangian~\cite{Bender:2005hf, Alexandre:2015oha, Alexandre:2015kra, Beygi:2019qab} and Yukawa theories~\cite{Alexandre:2015kra, Alexandre:2020bet}, with potential applications to flavor oscillations, e.g., in the quark or neutrino sector of the Standard Model~\cite{Jones-Smith:2009qeu, Ohlsson:2015xsa, Alexandre:2023afi}; and non-Hermitian supersymmetric quantum field theories~\cite{Bender:1997ps, Alexandre:2020wki}. Non-Hermitian quantum field theories that exhibit spontaneous symmetry breaking, the Goldstone theorem, and the Higgs mechanism have also attracted attention~\cite{Alexandre:2018uol, Mannheim:2018dur, Alexandre:2018xyy, Alexandre:2019jdb, Fring:2019hue, Fring:2019xgw, Fring:2020bvr}, as well as those permitting topological defects~\cite{Fring:2020xpi, Fring:2020wrj, Correa:2021pwi, Begun:2021wol, Correa:2021hfj}.

Many of these theories are constructed by appending non-Hermitian operators to an otherwise Hermitian Lagrangian, or by analytic continuation of Hermitian theories, for instance, by rotating coupling constants into the complex plane. Often times, these theories are analyzed in their $PT$-unbroken regimes (when the energy eigenvalues are real) by transforming to a Hermitian theory. However, if we instead consider the non-Hermitian theory directly, we run into issues with physical consistency, as detailed below.

In quantum field theory, the physical system is described by a Fock space $\fs$ consisting of multiparticle states, rather than a single-particle Hilbert space. The system dynamics is governed by a Hamiltonian operator $\hh:\fs\mapsto\fs$ acting on this Fock space. Similar to non-Hermitian quantum mechanics, if the Hamiltonian operator is pseudo-Hermitian $\hh^{\dag}=\heta \hh\heta^{-1}$ with respect to some Hermitian operator $\heta:\fs\mapsto\fs$, it will exhibit real eigenspectra, complex-conjugate pairs of eigenvalues, or exceptional points. As in pseudo-Hermitian quantum mechanics, we choose a new inner product $\braket{\cdot}{\cdot}_{\heta}\df \braket{\cdot}{\heta\cdot}$, instead of the Dirac inner product $\pd$, as the former yields real expectation values of the Hamiltonian.

In \comma\comma standard" quantum field theory, the Hamiltonian is a functional of field operators and their canonical momenta $(\hps,\hps^{\dag},\hp,\hp^{\dag})$, instead of position and momentum operators $(\hat{x},\hat{p})$, as it is in quantum mechanics. The time evolution of the field operators is governed by Hamilton's equations:
\be
 [\hps (\vx,t),\hh]=i\der{0}{}{\hps (\vx,t)} \text{\:\:\:and\:\:\:} [\hpsd (\vx,t), \hh^{\dag}]=i\der{0}{}{\hpsd (\vx,t)}\;.
\ee
Most notably, the field operator $\hps$ and its Hermitian conjugate $\hpsd$ do not evolve with the same Hamiltonian, since it is non-Hermitian $\hh^{\dag}\neq\hh$. This leads to mutual inconsistency in the Euler-Lagrange equations for $\hps$ and $\hpsd$, a common feature observed in various non-Hermitian quantum field theory models, first pointed out in Ref.~\cite{Alexandre:2017foi}. One option is to fix the dynamics with respect to only one of the Euler-Lagrange equations~\cite{Alexandre:2017foi}. For noninteracting theories, it can be argued that physical observables remain unchanged regardless of the chosen equation~\cite{Alexandre:2017foi}. However, in the context of interacting theories, this method leads to distinct physical results~\cite{Alexandre:2018uol, Alexandre:2018xyy, Alexandre:2019jdb} 
compared with approaches based on transforming the non-Hermitian theory to the Hermitian, i.e., standard quantum field theory~\cite{Mannheim:2018dur, Fring:2019hue, Fring:2019xgw, Fring:2020bvr}. It is also an open question as to how to consistently introduce gauge symmetries in non-Hermitian quantum field theory~\cite{Millington:2019dfn}. Moreover, it has been observed that the conserved currents in a non-Hermitian Yukawa theory are not invariant under proper Lorentz transformations~\cite{Alexandre:2020bet}. In this paper, we show that these inconsistencies naturally arise and are to be expected due to the quantum fields $\hps$ and $\hpsd$ not transforming in the same representation of spacetime symmetry transformations (the proper Poincar\'e group).  

Attempts have been made~\cite{Alexandre:2020gah, Alexandre:2022uns} to resolve these issues in some $PT$-symmetric models by redefining the conjugate field in terms of the parity transformation. In these models, the field and its parity conjugate evolve with the same Hamiltonian. In this work, we confirm that the conjugate field must be determined with care in order to build self-consistent pseudo-Hermitian quantum field theories. As was noted in Ref.~\cite{Chernodub:2021waz}, this requires us to consider non-Hermitian generators of the proper Poincar\'{e} group, which is clearly the case for the generator of time translations:\ the Hamiltonian $\hat{P}^{0}=\hh$. This marks a significant difference between pseudo-Hermitian quantum mechanics and pseudo-Hermitian quantum field theory, where, for the latter, spacetime symmetries play a pivotal role. Moreover, we will see that the non-Hermiticity of the generator of time translations implies that other generators are, in general, non-Hermitian. Using group transformation properties, we construct conjugate field operators transforming consistently under the full proper Poincar\'{e} group. 

The paper is organized as follows. In Sec.~\ref{sec:tt_invariance}, we reexamine time evolution in the case of pseudo-Hermitian quantum mechanics in terms of representations of time translations. In Sec.~\ref{sec:poincare_invariance}, we turn our attention to pseudo-Hermitian quantum field theory and, by demanding Poincar\'{e} invariance, we show that pseudo-Hermiticity of the Hamiltonian implies pseudo-Hermiticity of the remaining group generators. In Subsec.~3.1, we identify the relevant representations that act on the quantum field operators by considering their matrix elements with respect to the inner product $\braket{\cdot}{\heta\cdot}$. In Sec.~\ref{sec:Fock_space_reps}, we examine how quantum fields behave under spacetime translations and proper Lorentz transformations when Fock-space representations are non-Hermitian.

In Subsec.~4.2, we present the key result of this paper. Therein, we define the \comma\comma dual'' quantum field operator $\hat{\td{\psi}}^{\dag}$, which transforms in the dual representation of the proper Poincar\'e group of the field operator $\hps$. Hence, the Lagrangian composed of the field operator $\hps$ and its \comma\comma dual'' $\hat{\td{\psi}}^{\dag}$ transforms as one object in a single representation of the proper Poincar\'e group. This also leaves any bilinear combinations $\hat{\td{\psi}}^{\dag}\hps$ Poincar\'e invariant. Additionally, as the \comma\comma dual'' field operator evolves with the Hamiltonian $\hh$ instead of $\hh^{\dag}$, the Euler-Lagrange equations are automatically consistent, unlike those formulated in terms of the Hermitian conjugate field operator $\hpsd$.

In~Sec.~\ref{sec:finite_dim_reps}, we show how to construct pseudo-Hermitian finite-dimensional representations of the proper Lorentz group, which are necessary to define a \comma\comma dual'' field for fields higher than spin-$0$. We consider the specific cases of:\ spin-0 scalars, for which the finite-dimensional representations are trivial; spin-half Weyl spinors, which are the smallest nontrivial representation of the Lorentz Lie algebra; and spin-half Dirac fermions. Finally, in Sec.~\ref{sec:PT_example}, we apply our discussion to a specific example of a $PT$-symmetric 2-component complex scalar field theory, correctly determining the dual field and identifying the relevant discrete symmetries.


\section{Time-Translation Invariance in non-Hermitian Quantum Mechanics}
\label{sec:tt_invariance}

Before we turn to the case of non-Hermitian quantum field theory, it is helpful to first reexamine the concept of unitary time evolution in non-Hermitian quantum mechanics and its connection to representations of time translations. This will prove useful when considering representations in non-Hermitian quantum field theory.

As mentioned in the Introduction, the properties of real expectation values and unitary time evolution are not unique to Hermitian operators. In fact, a more general  condition for an observable $\hat{A}$ to yield real expectation values is for it to be \emph{pseudo-Hermitian} \cite{Mostafazadeh:2008pw}:
\begin{definition}[Pseudo-Hermitian]
\:\:\:\:\:\:\:\:\:\:\:\:\:\:\:\:\:\:\:\:\:\:\:\:\:\:\:\:\:\:\:\:\:\:\:\:\:\:\:\:\:\:\:\:\:\:\:\:\:\:\:\:\:\:\:\:\:\:\:\:\:\:\:\:\:\:\:\:\:\:\:\:\:\:\:\:\:\:\:\:\:\:\:\:\:\:\:\:\:\:\:\:\:\:\:\:\:\:\:\:\:\:\:\:\:\:\:\:\:\: 
An operator $\hat{A}:\hil\mapsto\hil$ is $\heta$-pseudo-Hermitian if and only if $\hat{A}^{\dag}=\heta\hat{A}\heta^{-1}$ with respect to some operator $\heta:\hil\mapsto\hil$ that is Hermitian $\heta^{\dag}=\heta$.
\end{definition} 
In contrast to a Hermitian Hamiltonian, an $\heta$-pseudo-Hermitian Hamiltonian allows for both real and complex energy eigenvalues~\cite{Mostafazadeh:2001jk}. In each case, the expectation value of an $\heta$-pseudo-Hermitian Hamiltonian is always real with respect to the inner product $\braket{\cdot}{\cdot}_{\heta}$. If the energy eigenvalues are real, then it is possible to find a similarity transformation that maps this non-Hermitian Hamiltonian to a Hermitian operator~\cite{Mostafazadeh:2002id}. However, the non-Hermitian Hamiltonian has no Hermitian counterpart at the exceptional points or when the eigenenergies
are complex. Hence, a pseudo-Hermitian Hamiltonian may describe a unique physical system.

When we say that an $\heta$-pseudo-Hermitian Hamiltonian has unitary time evolution, it is not exactly the same as the unitary time evolution in \comma`standard'' quantum mechanics. Instead, the time evolution is \comma`\emph{pseudounitary}'', which means it is unitary with respect to the operator $\heta$:

\begin{definition}[Pseudo-Unitary]
\:\:\:\:\:\:\:\:\:\:\:\:\:\:\:\:\:\:\:\:\:\:\:\:\:\:\:\:\:\:\:\:\:\:\:\:\:\:\:\:\:\:\:\:\:\:\:\:\:\: \:\:\:\:\:\:\:\:\:\:\:\:\:\:\:\:\:\:\:\:\:\:\:\:\: \:\:\:\:\:\:\:\:\:\:\:\:\:\:\:\:\:\:\:\:\:\:\:\:\: \:\:\:\:\:\:\:\:\:\:\:\:\:\:\:\:\:\:\:\:\:\:\:\:\:  An operator $\hu:\hil\mapsto\hil$ is $\heta$-pseudo-unitary if and only if $\hu^{\dag}\heta \hu=\heta$.    
\end{definition}

Indeed, in pseudo-Hermitian quantum mechanics, we can check that, given an $\heta$-pseudo-Hermitian Hamiltonian $\hh$, the time-evolution operator $\hu (t)=e^{-i\hh t}$ is $\heta$-pseudo-unitary. As a result, the time evolution of the \comma`ket'' states $\ket{\psi}\in\hil$ in our Hilbert space is governed by the $\heta$-pseudo-unitary operator $\hu(t)$:
\begin{equation}\ket{\psi (t)}=\hu (t)\ket{\psi (0)}=e^{-i\hh t}\ket{\psi (0)}\;.
\end{equation}
In the context of representations, the operator $\hu(t)=e^{-i\hh t}$ is the representation of time translations on the Hilbert space $\hil$, and the Hamiltonian $\hh$ is the generator 
of this representation.

The \comma\comma bra'' states $\bra{\psi}\in\hil^{*}$, however, are governed by $\hu(t)^{\dag}$:
\be
\bra{\psi(t)}=\bra{\psi(0)}\hu(t)^{\dag}=\bra{\psi(0)}e^{i\hh^{\dag}t}\:.
\ee
Hence, the dynamics of \comma\comma bra'' and \comma\comma ket'' states is not governed by the same Hamiltonian as it is non-Hermitian. They evolve under different representations of time translations.

Similarly, the wave function $\psi (\vx,t)\df\braket{\vx}{\psi (t)}\in\mathbb{F}(\hil)$ evolves with the Hamiltonian function $H:\mathbb{F}(\hil)\mapsto\mathbb{F}(\hil)$ acting on the vector space of wave functions $\mathbb{F}(\hil)$ of the Hilbert space $\hil$: 
\be
\psi (\vx,t)=e^{-i H t}\psi (\vx,0)\;.
\ee
However, the complex-conjugate wave function $\psi^{*}(\vx,t)\in\mathbb{F}(\hil^{*})$ evolves with the Hermitian conjugate $H^{\dag}:\fs (\hil^{*})\mapsto\fs (\hil^{*})$, where $\mathbb{F}(\hil^{*})$ is the vector space of wave functions of the dual Hilbert space $\hil^{*}$:
\be
\psi^{*} (\vx,t)=\psi^{*} (\vx,0)e^{i H^{\dag} t}\;.
\ee
As the Hamiltonian $H^{\dag}\neq H$ is non-Hermitian, the complex-conjugate wave function transforms in a different representation of time translations, generated by $H^{\dag}$ instead of $H$.

This has significant implications for the probability density, which is composed of two components:
\be
\mathbb{P} (\vx,t)\df\abs{\psi (\vx,t)}^{2}=\psi^{*} (\vx,t)\psi (\vx,t)\;,
\ee
viz.~the wave function $\psi$, evolving with the Hamiltonian $H$, and its complex conjugate $\psi^{*}$, evolving with $H^{\dag}$. Consequently, the overall object, i.e., the probability density, undergoes transformations in two distinct representations of time translations and is not conserved.

In pseudo-Hermitian quantum mechanics, this is fixed by defining the probability density with respect to the new inner product $\braket{\cdot}{\heta\cdot}$:
\be
\braket{\psi(t)}{\heta\:\psi(t)}=\int {\rm d}^{3}x\: \td{\mathbb{P}}(\vx,t)=\braket{\psi(0)}{\heta\:\psi(0)}\;.
\ee
This inner product remains invariant under time translations generated by a $\heta$-pseudo-Hermitian Hamiltonian. Thus, the probability density is conserved.

However, another interpretation of this result, which will be relevant in the next section, is to introduce a new object called the \comma`dual'' wave function $\td{\psi}^{*}$, which evolves with the same Hamiltonian as $\psi$. This enables us to redefine the probability density in terms of the wavefunction $\psi$ and its \comma\comma dual'' $\td{\psi}^{*}$, so that the probability density transforms in a single representation of time translations. 

To find the dual wave function, we consider $(U,\hil)$ to be a representation of time translations on the Hilbert space $\hil$. We define a \comma`dual'' representation $(U^{*},\hil^{*})$ on the dual Hilbert space $\hil^{*}$ as
\begin{definition}[Dual Representation]
\be
\begin{aligned}
    \hu^{*}(t):\: &\hil^{*} \longmapsto \hil^{*}\\
    &\bra{\td{\phi}} \longmapsto \hu^{*}(t)[\bra{\td{\phi}}]\df\bra{\td{\phi}}\hu^{-1}(t)
\end{aligned}
\ee 
\end{definition}
\noindent Here, $\bra{\td{\phi}}\in\hil^{*}$ are the \comma`dual'' states, i.e., the states that transform in the dual representation:
\begin{definition}[Dual States]
\be
\begin{aligned}
\bra{\td{\phi}}:\:&\hil \longmapsto \cn \\
&\ket{\psi}\longmapsto \braket{\td{\phi}}{\psi}
\end{aligned}
\ee
\end{definition}
\noindent For an $\heta$-pseudo-unitary representation $\big(\hu,\hil\big)$, the dual states are related to the bra states by $\bra{\td{\phi}}=\bra{\phi}\heta$. The conjugate representation $\big(\hu^{\dag},\hil^{*}\big)\cong\big(\hu^{*},\hil^{*}\big)$ is isomorphic to the dual representation, as they are related by the similarity transformation $\hu^{\dag}=\heta\hu^{-1}\heta^{-1}$, since $\hu$ is $\heta$-pseudo-Hermitian.

We can now define the dual wave functions $\td{\psi}^{*}:\mathbb{F}(\hil^{*})\mapsto\mathbb{F}(\hil^{*})$, as the wave function transforming in the dual representation of $\psi:\mathbb{F}(\hil)\mapsto\mathbb{F}(\hil)$:

\begin{definition}[Dual Wavefunction]
 \be
\td{\psi}^{*} (\vx,t)\df \langle \td{\psi}(t)\mid\vx\rangle=\bra{\psi (t)}\heta\ket{\vx}=\bra{\psi (0)}\underbrace{e^{i\hh^{\dag}t}\heta}_{\mathclap{\heta\text{-pseudo-Hermitian}}}\ket{\vx}=\bra{\psi (0)}\heta e^{i\hh t}\ket{\vx}=\bra{\psi (0)}\heta\ket{\vx}e^{i H t}\;.
\ee   
\end{definition}
\noindent Hence, we see that its time evolution is governed by $H$ and not $H^{\dag}$.

For the special case where $\heta$ is itself a coordinate transformation (e.g., parity), its action on the position eigenstates can be described as follows:
\be
\heta\ket{\vx}=\eta \ket{\vx_{\eta}}, \text{\:\:\: where\:\:} \eta\in\cn \text{\:\:is the phase.}
\ee
The dual wave function simplifies to
\be
\td{\psi}^{*} (\vx,t)=\td{\psi}^{*} (\vx,0) e^{i H t}\text{\:\:\:where\:\:\:}\td{\psi}^{*} (\vx,0)=\psi^{*} (\vx_{\eta},0)\eta \;,
\ee

Finally, we define the probability density in terms of the wave function $\psi$ and its dual $\td{\psi}^{*}$:
\be
\tilde{\mathbb{P}} (\vx,t)\df\td{\psi}^{*} (\vx,t)\psi (\vx,t)=\td{\psi}^{*} (\vx,0)\psi (\vx,0)\;,
\ee
which transforms in a single representation generated by $H$ and is now conserved.

The above procedure is equivalent to defining a new inner product $\braket{\cdot}{\heta \cdot}$ on the Hilbert space $\hil$:
\be
\braket{\psi (t)}{\heta \psi (t)}=\int {\rm d}^{3}x \:\bra{\psi (t)}\heta\ket{\vx}\braket{\vx}{\psi (t)}=\int {\rm d}^{3}x\:\td{\mathbb{P}} (\vx,t),
\ee
which is invariant under time translations.

The above considerations are nonrelativistic. In the relativistic case of quantum field theory, we are not only concerned with time translation invariance, but also with Lorentz invariance. Specifically, the symmetry group of Minkowskian quantum field theory is the proper Poincar\'e group I$\SOP=\SOP\rtimes \rn^{1,3}$. It is  composed of the proper Lorentz transformations $\SOP$ and spacetime translations $\rn^{1,3}$. Instead of position and momentum operators, we have quantum field and canonical momentum field operators, and, as we will show in the next section, we face similar issues when trying to find conjugate field operators that transform correctly in the dual representation of proper Poincar\'e transformations.


\section{Poincar\'e Invariance in non-Hermitian quantum field theory}\label{sec:poincare_invariance}

We now turn our attention to quantum field theory. In the canonical operator formulation, the Hamiltonian operator $\hh:\fs\mapsto\fs$ acts on the Fock space $\fs$. It is a function of field operators and their canonical momenta, i.e., $\hh=\hh\big(\hps,\hpsd,\hp,\hp^{\dag}\big)$. Through the rest of the paper, we assume that the Hamiltonian $\hh^{\dag}=\heta\hh\heta^{-1}$ is $\heta$-pseudo-Hermitian with respect to some Hermitian operator $\heta:\fs\mapsto\fs.$ 

The \comma`ket'' states in Fock space $\ket{\al (t)}\in\fs$ evolve subject to the Schr\"{o}dinger equation
\be
i\partial_{t}\ket{\al (t)}=\hh\ket{\al (t)}\;.
\ee
Just as in pseudo-Hermitian quantum mechanics, their time evolution is $\heta$-pseudo-unitary. We define our Fock space $\fs=(\indices{^{\otimes}}\hil,\pd_{\heta})$, with respect to the inner product $\braket{\cdot}{\cdot}_{\heta}=\braket{\cdot}{\heta\cdot}$, which is invariant under time translations and yields real energy expectation values.

The time evolution of the field operators $\hps$ and $\hpsd$ is governed by Hamilton's equations of motion. However, by considering the Hermitian conjugate of the Hamilton's equation for $\hps$, namely,
\be\label{eq:Hamiltons_eqs}
 [\hps (\vx,t),\hh]=i\der{0}{}{\hps (\vx,t)}\implies  [\hpsd (\vx,t),\hh^{\dag}]=i\der{0}{}{\hpsd (\vx,t)}\;,
\ee
we observe that while $\hps$ evolves with the Hamiltonian $\hh$, its Hermitian conjugate $\hpsd$ evolves with $\hh^{\dag}$. As the Hamiltonian is non-Hermitian, this implies that the two fields are subject to different Hamiltonians. We also see this in the Heisenberg picture:
\be\label{eq:Heisenber_pic}
\hps (\vx,t)=e^{i\hh t}\hps (\vx)e^{-i\hh t}\text{\:\:\:and\:\:\:}
\hpsd (\vx,t)=e^{i\hh^{\dag} t}\hpsd (\vx)e^{-i\hh^{\dag} t}\;.
\ee
In the language of representations, the conjugate field $\hpsd$ does not transform in the dual representation of $\hps$. This implies that the kinetic and mass terms are not invariant under time translations, namely,
\be\label{eq:bilinear_combo}
\hpsd  (\vx,t)\hps (\vx,t)=e^{i\hh^{\dag} t}\hpsd (\vx)e^{-i\hh^{\dag} t}e^{i\hh t}\hps (\vx)e^{-i\hh t}
\neq e^{i\hh  t}\hpsd (\vx)\hps (\vx)e^{-i\hh t}\;.
\ee

Our aim is to find the \comma`dual'' field operator that transforms in the dual representation. However, quantum field theory in Minkowski spacetime has to be invariant under proper Poincar\'{e} transformations I$\SOP$, where time and space transformations are intermixed. The Hamiltonian $\hh$ is the generator of time translations, but the non-Hermiticity of $\hh$ will turn out to translate into the non-Hermiticity of other generators of the proper Poincar\'e group. This observation is the primary focus of this section.

Let $\big(\hu,\fs\big)$ be a $\heta$-pseudo-unitary representation of I$\SOP$ on the Fock space $\fs$. Any element of the proper Poincar\'{e} group can be expanded in terms of generators of the Poincar\'e Lie algebra $\mathfrak{i}\so$, i.e.,
\be
\hu (\ep,\lt)=e^{\frac{i}{2}\omega_{\mu\nu}\hJ^{\mu\nu}}e^{i\ep_{\mu}\hpm^{\mu}}=\hat{\id}+\frac{i}{2}\omega_{\mu\nu}\hJ^{\mu\nu}+i\ep_{\mu}\hpm^{\mu}\cdots\;.
\ee
Herein, $\hJ^{0 i}$ are the generators of boosts, $\hJ^{ij}$ are the generators of rotations, the Hamiltonian $\hpm^{0}=\hh$ is the generator of time translations, and the $3$-momentum operator $\hpm^{i}$ is the generator of space translations.

Now, if the inner product $\braket{\cdot}{\heta\cdot}$ is invariant under proper Poincar\'e transformations then the representations of I$\SOP$ are $\heta$-pseudo-unitary: $\hu^{\dag}\heta\hu=\heta$ for all group elements $
 (\ep,\lt)\in$ I$\SOP$. Expanding $\hu$ implies that the generators are $\heta$-pseudo-Hermitian, i.e.,
\be\label{eq:non_Hermiticity_of_other_generators}
\hJ^{\mu\nu \dag}=\heta\hJ^{\mu\nu}\heta^{-1},\:\:\:\hpm^{\mu\dag}=\heta\hpm^{\mu}\heta^{-1}  .
\ee
This can be seen directly by considering the Poincar\'e Lie algebra
\begin{align}\label{eq:Poincare_Lie_Braket}
\left[\hJ^{\mu \nu}, \hJ^{\rho \sigma}\right] & =i (g^{\mu \sigma} \hJ^{\nu \rho}+g^{\nu \rho} \hJ^{\mu \sigma}-g^{\mu \rho} \hJ^{\nu \sigma}-g^{\nu \sigma} \hJ^{\mu \rho}) \nonumber\\
\left[\hpm^\mu, \hJ^{\rho \sigma}\right] & =i (g^{\mu \rho} \hpm^\sigma-g^{\mu \sigma} \hpm^\rho) \nonumber\\
{\left[\hpm^\mu, \hpm^\nu\right] } & =0\;,
\end{align}
where $g^{\mu\nu}=\text{diag} \big(1,-1,-1,-1\big)$ is the Minkowski metric. Taking $\mu=0$ in the second bracket, we have
\be
 [\hh,\hJ^{0 i}]=i\hpm^{i}\implies
 [\heta\hh\heta^{-1},\hJ^{0 i \dag}]=i\hpm^{i\dag}\implies  [\hh,\heta^{-1}\hJ^{0 i \dag}\heta]=i\heta^{-1}\hpm^{i\dag}\heta\;,
\ee
which also implies that the generators cannot be Hermitian, unless they commute with $\heta$.

\subsection{Connection to classical fields}\label{subsec:connection_to_classical_fields}

It is important to consider the connection between quantum and classical fields, at the very least to identify the relevant representations for our subsequent discussions.

Let us consider an $n$-component quantum field $\hps^{a}(x),\:a\in\{1,\cdots,n\}$. In Hermitian quantum field theory, the matrix elements of this field would be defined as an $n$-component function using the Dirac inner product $\pd$:
\be
\mathcal{M}^{a}_{\al\bt}(x)=\bra{\al}\hps^{a} (x)\ket{\bt}
\ee
for given Fock-space states $\ket{\al},\ket{\bt}\in\fs$.

However, in the case of a non-Hermitian Hamiltonian, the energy eigenstates are not orthogonal with respect to the Dirac inner product $\pd$. Instead, if the Hamiltonian is $\heta$-pseudo-Hermitian, the eigenstates with real eigenvalues become orthogonal with respect to the inner product $\braket{\cdot}{\heta\cdot}$. As a result, the matrix elements are defined in terms of this inner product as follows:
\be
\td{\mathcal{M}}^{a}_{\al\bt}(x)\df\braket{\al}{\hps^{a}(x)\bt}_{\heta}=\bra{\al}\heta\:\hps^{a}(x)\ket{\bt}.
\ee
In this way, we define the expectation value of the quantum field operator as
\be\label{eq:exp_val_heta}
\Psi (x)\df\bra{\al}\hps (x)\ket{\al}_{\heta}\:.
\ee
Note that the operator $\heta$ and the field operator $\hps^{a}(x)$ do not, in general, commute. 

By postulating that the expectation values of quantum fields possess the same transformation properties as classical field functions under the Poincar\'e group, we establish a connection between the transformation properties of quantum fields and their classical counterparts. In the quantum field theory literature, this is known as the \comma`correspondence principle''~\cite{Greiner}, \cite{Bogoliubov}.

A typical element $(\ep, \lt)\in\text{I}\SOP$ of the proper Poincar\'e group is composed of a translation by a constant $4$-vector $\ep \in \mathbb{R}^{1,3}$ and a proper Lorentz transformation $\lt \in \SOP$. Under these transformations, the expectation value of the quantum field operator undergoes changes in three distinct representations of the proper Poincaré group:

\begin{enumerate}
    \item The infinite-dimensional \comma`\emph{coordinate representation}'' acting on the spacetime coordinates:
    \be
    \begin{aligned}
        (\ep,\lt):\:\:\: &\rn^{1,3}\longmapsto \:\:\rn^{1,3}\\
        &x\:\:\:\:\:\longmapsto \:\:(\ep,\lt)[x]=\lt \cdot x+\ep\;.
    \end{aligned}
    \ee
    
    \item The infinite-dimensional \comma`\emph{Fock-space representation}'' $\big(\hu,\fs\big)$, acting on states in Fock space:
\be
\begin{aligned}
\hu(\ep,\lt):\:\:\:&\fs\:\:\longmapsto\fs\\
    &\ket{\al}\longmapsto \hu(\ep,\lt)\ket{\al}\;.
\end{aligned}
\ee

\item The \comma`\emph{finite-dimensional representation}'' $ (D,\cn^{n})$ of the proper Lorentz group $\SOP$, mixing components of an $n$-component field:
\be
\begin{aligned}
    D(\lt):\:\:\:&\cn^{n}\longmapsto \cn^{n}\\
    &\hps^{a} \longmapsto D\indices{^a_b} (\lt)\hps^{b}\;,
\end{aligned}
\ee
where $a,b\in {1,\cdots,n}$.

\end{enumerate}

\section{Fock-Space Representations of the Proper Poincar\'e Group}\label{sec:Fock_space_reps}

In the previous section, we showed that the conjugate field $\hpsd$ evolves with a different Hamiltonian to $\hps$. As a result, the bilinear combination $\hpsd\hps$ in the Lagrangian is not invariant under time translations~\eqref{eq:bilinear_combo}. Moreover, if we look at the full proper Poincar\'e group I$\SOP$, we see that the non-Hermiticity of the Hamiltonian implies non-Hermiticity of other generators~\eqref{eq:non_Hermiticity_of_other_generators}, unless they commute with the Hermitian operator $\heta$. Thus, if we wish to construct a quantum field theory that is invariant under I$\SOP$, we cannot construct it from $\hps$ and $\hpsd$. Instead, we need to find a new quantum field operator $\hat{\td{\psi}}^{\dag}$ that transforms in the \comma\comma dual'' representation of $\hps$, just as we did with the \comma\comma dual'' wave function in Sec.~\ref{sec:tt_invariance}. 

To find the \comma\comma dual'' field operator, we consider $(\hu,\fs)$ to be a representation of the proper Poincar\'e group I$\SOP$ on the Fock space $\fs$. We define a \comma\comma dual'' representation $(\hu^{*},\fs^{*})$ on the \comma\comma dual'' Fock space $\fs^{*}$ as
\begin{definition}[Dual Fock-Space Representation]
\be
\begin{aligned}
    \hu^{*}(\ep,\lt):\: &\fs^{*} \longmapsto \fs^{*}\\
    &\bra{\td{\al}} \longmapsto \hu^{*}(\ep,\lt)[\bra{\td{\al}}]\df\bra{\td{\al}}\hu^{-1}(\ep,\lt)\:.
\end{aligned}
\ee 
\end{definition}
\noindent Here, $\bra{\td{\al}}\in\fs^{*}$ are the \comma`dual'' states, i.e., the states that transform in the \comma\comma dual'' Fock-space representation. For an $\heta$-pseudo-unitary representation $(\hu,\fs)$, the \comma\comma dual'' states are related to the \comma\comma bra'' states by $\bra{\td{\al}}=\bra{\al}\heta$.

However, as noted in the previous section, if the quantum field $\hps^{a}$ is multicomponent, its components will mix under finite-dimensional representations $(D,\cn^{n})$ of the proper Lorentz group $\SOP$. We define a \comma\comma dual'' representation $(D^{*},\cn^{n *})$ on the \comma\comma dual'' vector space $\cn^{n *}$ as
\begin{definition}[Dual Finite-Dimensional Representation]
\be
\begin{aligned}
    D^{*}(\lt):\: &\cn^{n *} \longmapsto \cn^{n *}\\
    & \hat{\td{\psi}}^{\dag a} \longmapsto D^{*}(\lt)[\hat{\td{\psi}}^{\dag a}]\df\hat{\td{\psi}}^{\dag b}D\indices{^{-1}_b^a}(\lt)\:.
\end{aligned}
\ee 
\end{definition}

Given that the quantum field operator $\hps$ transforms in a Fock-space representation $(\hu,\fs)$ and a finite-dimensional representation $(D,\cn^{n})$, we define the \comma\comma dual'' field operator $\hat{\td{\psi}}^{\dag}$ as the field operator that transforms in a \comma`dual'' Fock-space representation $(\hu^{*},\fs^{*})$ and a \comma\comma dual'' finite-dimensional representation $(D^{*},\cn^{n *})$. We then say that the quantum field $\hat{\td{\psi}}^{\dag}$ transforms in the \comma\comma dual'' representation of $\hps$.

We postulate that, under the proper Poincar\'e group I$\SOP$, the expectation value of an $n$-component quantum field~\eqref{eq:exp_val_heta} transforms as an $n$-component classical function:
\be
\begin{aligned}
\text{I}\SOP:\: &\:\:\:
x\:\:\:\:\:\longmapsto\:\:\: x'=\lt\cdot x+\ep\:,\\
    &\Psi^{a}(x) \longmapsto \Psi^{a '}(x')=D\indices{^a_b}(\lt)\Psi^{b}(x)\:,\\
    &\td{\Psi}^{\dag a}(x) \longmapsto \td{\Psi}^{\dag a '}(x')=\td{\Psi}^{\dag b}(x)D\indices{^{-1}_b^a}(\lt)\:.
\end{aligned}
\ee
\noindent Here, $\Psi^{a}(x)\df \bra{\td{\al}}\hps^{a}(x)\ket{\al}$ is the expectation value of the quantum field $\hps$ transforming in the finite-dimensional representation $(D,\cn^{n})$, while $\td{\Psi}^{\dag a}(x)\df \bra{\td{\al}}\thps{\dag a}{x}\ket{\al}$ is the expectation value of the \comma\comma dual'' quantum field $\hat{\td{\psi}}^{\dag}$ transforming in the \comma\comma dual'' finite-dimensional representation $(D^{*},\cn^{n})$.

Considering that the states $\ket{\al}\in\fs$ transform under the Fock-space representation $(\hu,\fs)$ and their \comma\comma dual'' states $\bra{\td{\al}}\in\fs^{*}$ under the \comma\comma dual'' Fock-space representation $(\hu^{*},\fs^{*})$:
\be
\begin{aligned}
\text{I}\SOP: \:&\bra{\td{\al}}\hps^{a}(x)\ket{\al} \:\longmapsto \bra{\td{\al}}\hu^{-1}\hps^{ a}(x')\hu\ket{\al}=D\indices{^a_b}\bra{\td{\al}}\hps^{b}(x)\ket{\al}\:,\\
&\bra{\td{\al}}\thps{\dag a}{x}\ket{\al} \longmapsto \bra{\td{\al}}\hu^{-1}\thps{\dag a}{x'}\hu\ket{\al}=\bra{\td{\al}}\thps{\dag b}{x}\ket{\al}D\indices{^{-1}_b^a}\:,
\end{aligned}
\ee
we derive the transformation laws for the quantum field $\hps$ and its \comma\comma dual'' $\hat{\td{\psi}}^{\dag}$:
\begin{align}
\hu^{-1}(\ep,\lt) \hps^{a}(x) \hu (\ep,\lt)&=D\indices{^a_b}(\lt)\hps^{b}(\lt^{-1}(x-\ep))\:,\label{eq:transf_of_quantum_field}\\
\hu^{-1}(\ep,\lt)\thps{\dag a}{x}\hu (\ep,\lt)&=\thps{\dag b}{\lt^{-1}(x-\ep)}D\indices{^{-1}_b^a}(\lt)\:.\label{eq:transf_of_dual_field}
\end{align}

Taking the Hermitian conjugate of Eq.~\eqref{eq:transf_of_quantum_field}, we obtain the transformation law for the conjugate field $\hpsd$:
\be\label{eq:transf_law_conj_field}
\hu^{\dag}(\ep,\lt)\hat{\psi}^{\dag a} (x)\hu^{-1 \dag}(\ep,\lt)=\hat{\psi}^{\dag b}(\lt^{-1}(x-\ep))D\indices{^\dag_b^a}(\lt)\:,
\ee
which clearly shows that the Hermitian-conjugate field $\hpsd$ does not transform in the \comma\comma dual'' representation of $\hps$, unless $\hu$ and $D$ are both unitary representations, i.e., the Hamiltonian $\hh$ is Hermitian. In fact, in Subsec.~4.2, we demonstrate that nonunitary Fock-space representations $(\hu,\fs)$ of the proper Lorentz group $\SOP$ imply nonunitarity of the finite-dimensional representations $(D,\cn^{n})$ and vice versa. For now, let us assume that the Fock-space representation is $\heta$-pseudo-unitary $\hu^{\dag}\heta\hu=\heta$, which is the case for an $\heta$-pseudo-Hermitian Hamiltonian $\hh$. We also assume that the finite-dimensional representation is $\pi$-pseudo-unitary $D^{\dag}\pi D=\pi$ with respect to some $n\times n$ Hermitian matrix $\pi:\cn^{n}\mapsto\cn^{n}$. Using this, we rearrange Eq.~\eqref{eq:transf_law_conj_field}:
\be\label{eq:transf_law_conj_dual}
\hu^{-1}(\ep,\lt)\left[ \heta^{-1} \hps^{\dag a}(x)\heta\: \pi\right] \hu (\ep,\lt) =\left[\heta^{-1}\hps^{\dag b}(\lt^{-1}(x-\ep))\heta\: \pi\right] D\indices{^{-1}_b^a}(\lt)\:.
\ee
This is exactly the transformation law of the \comma\comma dual'' field~\eqref{eq:transf_of_dual_field}. Hence, in general, the \comma\comma dual'' field operator will be of the form
\be\label{eq:main_result_1}
    \thps{\dag}{x}\df \heta^{-1}\hps^{\dag} (x^\eta)\heta\: \pi\:.
\ee
Since $\heta:\fs\mapsto\fs$ can in general be a coordinate transformation (e.g., parity), we need to include its action on the coordinates  $x^{\eta}$ (e.g., $x^{P}$) in the definition of the \comma\comma dual'' field. The field~\eqref{eq:main_result_1} transforms in the \comma\comma dual'' representation of the proper Poincar\'e group. As we did not assume anything about the spin of the quantum field $\hps$, the definition in~\eqref{eq:main_result_1} holds for fields of any spin. Thus, we will use it to define the \comma\comma dual'' fields for non-Hermitian scalar and fermionic quantum field theories in Secs.~\ref{sec:finite_dim_reps} and~\ref{sec:PT_example}.

In the following subsections, we examine how quantum fields behave under the generators of spacetime translations and proper Lorentz transformations. In particular, in Subsec.~4.1, we derive Hamilton's equations, confirming that they are inconsistent for the quantum field $\hps$ and its Hermitian conjugate $\hpsd$, but are in agreement with the \comma\comma dual'' field $\hat{\td{\psi}}^{\dag}$ defined above. In Subsec.~4.2, we show that non-Hermiticity of the Fock-space generators, originating from the non-Hermitian Hamiltonian $\hh$, directly leads to non-Hermitian generators of finite-dimensional representations of the proper Lorentz group $\SOP$. 


\subsection{Spacetime translations}\label{sec:spacetime_translations}

Consider a translation by a constant $4$-vector $x\mapsto x'=x+\ep$. The corresponding transformation of a quantum field $\hps^{a}$, according to Eq.~\eqref{eq:transf_of_quantum_field}, is
\be\label{eq:spacetime_translations_transformation_law}
\hu^{-1}(\ep)\hps^{a}(x)\hu (\ep)=\hps^{a}(x-\ep)\;.
\ee
Here, $\hu(\ep)=\hu(\ep,\id)$ for some constant $4$-vector $\ep\in\rn^{1,3}$. 

Since the components of a multicomponent field do not mix under spacetime translations, the only representations acting on the quantum field are the Fock-space and coordinate representations. Both of these can be expanded in terms of their generators:
\be
\hu (\ep)=e^{i\ep_{\mu}\hat{P}^{\mu}}\text{\:\:\:and\:\:\:} \hps^{a}(x-\ep)=e^{-\ep^{\mu}\partial_{\mu}}\hps^{a}(x)\:,
\ee
where $\hpm^{\mu}$ are the four generators of spacetime translations in the Fock-space representation and $\partial_{\mu}$ are the generators of spacetime translations in the coordinate representation.

Expanding each side in Eq.~\eqref{eq:spacetime_translations_transformation_law} gives us the relationship between the generators of spacetime translations for Fock-space and coordinate 
representations:
\be\label{eq:spacetimegenerators_psi}
 [\hps (x),\hat{P}_{\mu}]=i\partial_{\mu}\hps (x)\;.
\ee
Notably, for $\mu=0$, we recover Hamilton's equation of motion:
\be
\label{eq:HamiltonsEq}
 [\hps (x),\hh]=i\der{0}{}{\hps (x)}.
\ee

Taking a Hermitian conjugate of the above, we find that the conjugate field $\hpsd$ evolves with Hermitian conjugates of these generators:
\be
 [\hpsd (x),\hpm_{\mu}^ {\dag}]=i\der{\mu}{}{\hpsd (x)}\;,
\ee
In particular, for $\mu=0$, the conjugate field $\hpsd$ evolves with $\hh^{\dag}$ instead of $\hh$:
\be
 [\hpsd (x),\hh^{\dag}]=i\der{0}{}{\hpsd (x)}\;.
\ee

In Sec.~\ref{sec:poincare_invariance}, we showed that if the Hamiltonian is $\heta$-pseudo-Hermitian, then the 3-momentum operator $\hpm^{i}$ is also $\heta$-pseudo-Hermitian, i.e., $\hpm^{i\dag}=\heta \hpm^{i}\heta^{-1}$ for $i=1,2,3$. Hence, it is Hermitian if and only if it commutes with $\heta$. Using this, we rearrange the commutator:
\be
[\heta^{-1}\hps^{\dag} (x)\heta,\hat{P}_{\mu}]=i\partial_{\mu} (\heta^{-1}\hps^{\dag} (x)\heta)\;.
\ee
We see that the \comma dual' field defined by Eq.~\eqref{eq:main_result_1} will evolve with the same set of generators of spacetime translations as the quantum field $\hps$:
\be
 [\thps{\dag}{x},\hat{P}_{\mu}]=i\partial_{\mu}\thps{\dag}{x}\;.
\ee
In particular, the \comma dual' field $\hat{\td{\psi}}^{\dag}$ evolves with the same Hamiltonian $\hh$:
\be
 [\thps{\dag}{x},\hh]=i\der{0}{}{\thps{\dag}{x}}\;,
\ee
Thus, a Lagrangian composed of the \comma\comma dual'' field $\hat{\td{\psi}}^{\dag}$ and the quantum field $\hps$ will yield consistent equations of motion, and the bilinear terms $\hat{\td{\psi}}^{\dag}\hps$ will be invariant under spacetime translations.


\subsection{Proper Lorentz transformations}\label{sec:Lorentz_transformations}

Consider a proper Lorentz transformation $x\mapsto x'=\lt\cdot x$. Unlike spacetime translations, proper Lorentz transformations mix the components of multicomponent fields (both classical and quantum). Hence, the corresponding transformation of an $n$-component quantum field $\hps^{a}$ according to Eq.~\eqref{eq:transf_of_quantum_field} is
\be
 \hu^{-1}(\lt)\hps^{a}(x)\hu (x)=D\indices{^a_b}(\lt)\hps^{b}(\lt^{-1}x)\:.
\ee
Here, $\hu (\lt)=\hu (0,\lt)$, for some proper Lorentz transformation $\lt\in\SOP$. The mixing of field components is described by an $n\times n$ matrix $D (\lt)$, given by an $n$-dimensional matrix representation of the proper Lorentz group $\SOP$.

The representations acting on a quantum field are the Fock-space, finite-dimensional and coordinate representations. All of these can be expanded in terms of their generators:
\be\label{eq:exponential_finite_dim_reps}
\hu (\lt)=e^{\frac{i}{2}\omega_{\mu\nu}\hJ^{\mu\nu}},\:\:\:
D\indices{^a_b} (\lt)=e^{\frac{i}{2}\omega_{\mu\nu} (M^{\mu \nu})\indices{^a_b}}\text{\:\:\:and\:\:\:} \hps^{a}(\lt^{-1}x)=e^{-\frac{1}{2}\omega_{\mu\nu}m^{\mu\nu}}\hps^{a}(x)\:,
\ee
where $\hJ^{\mu\nu}$ are the six generators of rotations and boosts in the Fock-space representation, $M^{\mu\nu}$ are the $n\times n$ matrix generators of rotations and boosts in the $n$-dimensional matrix representation, and $m_{\mu \nu}=x_{\mu}\partial_{\nu}-x_{\nu}\partial_{\mu}$ are the generators of rotations and boosts in the coordinate representation.

Expanding each side in Eq~\eqref{eq:exponential_finite_dim_reps} gives us the relationship between the generators of proper Lorentz transformations $\SOP$ for the Fock-space, finite-dimensional and coordinate representations:
\be \label{eq:51}
 [\hps^{a} (x),\hJ^{\mu\nu}]= ( (M^{\mu\nu})\indices{^a_b}+im^{\mu\nu}\delta\indices{^a_b})\hps^{b} (x)\;.
\ee

Taking a Hermitian conjugate of the above equation, we find that the conjugate field $\hpsd$ evolves with Hermitian conjugates of these generators:
\be \label{eq:1}
 [\hps^{\dag a} (x),\hJ^{\dag \mu\nu}]=\hps^{\dag b} (x) (- (M^{\dag \mu \nu})\indices{_b^a}+im^{\mu \nu}\delta\indices{_b^a})\;,
\ee
For a unitary finite-dimensional representation $(D,\cn^{n})$, the generators $M^{\mu\nu\dag}=M^{\mu\nu}$ are Hermitian. This would imply that the Fock-space generators $\hJ^{\mu\nu \dag}=\hJ^{\mu\nu}$ are also Hermitian, meaning the Fock-space representation $(\hu,\fs)$ is unitary. However, according to Sec.~\ref{sec:poincare_invariance}, if the Hamiltonian operator $\hh$ is $\heta$-pseudo-Hermitian, i.e., $\hh^{\dag}=\heta\hh\heta^{-1}$, then other generators of the proper Lorentz transformations are also $\heta$-pseudo-Hermitian: $\hJ^{\mu\nu \dag}=\heta \hJ^{\mu\nu}\heta^{-1}$. Hence, the Fock-space generators $\hJ^{\mu\nu}$ are Hermitian if and only if they commute with $\heta$. Thus, if the Fock-space generators are not Hermitian, neither are the generators of finite-dimensional representations and vice versa. 

Using the pseudo-Hermiticity of the Fock-space generators, we can rewrite Eq.~\eqref{eq:1} as
\be
[\heta^{-1}\hps^{\dag} (x)\heta,\hJ^{ \mu\nu}]=\heta^{-1}\hps^{\dag} (x)\heta (-M^{\dag\mu\nu}+im^{\mu\nu})\;.
\ee
We can use a biorthonormal basis~\cite{Mostafazadeh:2001jk,Mostafazadeh:2008pw} to construct a Hermitian $n\times n$ matrix $\pi:\cn^{n}\mapsto \cn^{n}$ such that the generators $M^{\mu\nu \dag}=\pi M^{\mu\nu}\pi^{-1}$ are $\pi$-pseudo-Hermitian. This means that the finite-dimensional representation $D^{\dag}\pi D=\pi$ is $\pi$-pseudo-unitary. In the next section, we will show how to construct pseudo-Hermitian finite-dimensional representations from the representation theory of $\SOP$.

Further, applying the pseudo-Hermiticity of matrix generators, we find the field which evolves with the same set of generators as the field operator $\hps$:
\be
[\heta^{-1}\hps^{\dag} (x)\heta\:\pi,\hJ^{ \mu\nu}]=\heta^{-1}\hps^{\dag} (x)\heta\:\pi (-M^{\mu\nu}+im^{\mu\nu})\;.
\ee
This is exactly the \comma\comma dual'' field defined in Eq.~\eqref{eq:main_result_1}. Thus, the \comma\comma dual'' field $\hat{\td{\psi}}^{\dag}$ evolves with the same generators of proper Lorentz transformations as $\hps$ in Eq.~\eqref{eq:51}:
\be
 [\thps{\dag a}{x},\hJ^{\mu\nu}]=\thps{\dag b}{x} (- (M^{\mu\nu})\indices{_b^a}+im^{\mu\nu}\delta\indices{_b^a})\;.
\ee
Hence, bilinear operators of the form
$\hat{\td{\psi}}^{\dag}\hps$ will be invariant under the proper Lorentz transformations.


\section{Pseudo-Hermitian Finite Dimensional Representations}\label{sec:finite_dim_reps}

If we consider a multicomponent quantum field $\hps^{a}$ (e.g., a fermion field), its components will mix under proper Lorentz transformations $\SOP$. The action of the proper Lorentz group on an $n$-component field is given by an $n$-dimensional matrix representation $(D, \mathbb{C}^n)$:
\be
\begin{aligned}
D(\lt):\:\:\:&\cn^{n}\longmapsto \cn^{n}\\
    &\hps^{a} \longmapsto D\indices{^a_b}(\lt)\hps^{b}\;.
\end{aligned}
\ee
\noindent Here, $D(\lt)$ is an $n\times n$ matrix determined by a proper Lorentz transformation $\Lambda \in \SOP$.

In Eq.~\eqref{eq:1}, we noted that if the generators of the Fock-space representation are non-Hermitian $\hJ^{\dag \mu\nu}\neq\hJ^{\mu\nu}$, then the generators of finite-dimensional representations are not, in general, Hermitian either $M^{\mu\nu^\dag}\neq M^{\mu\nu}$. Nonetheless, if we can find a Hermitian $n\times n$ matrix $\pi:\cn^{n}\mapsto\cn^{n}$, such that these generators are $\pi$-pseudo-Hermitian $M^{\mu\nu\dag}=\pi M^{\mu\nu}\pi^{-1}$, then we can derive the dual quantum field $\hat{\td{\psi}}^{\dag}$, which transforms in the dual representation of the quantum field $\hps$.

In this section, we explore how the pseudo-Hermitian finite-dimensional representations naturally emerge in the representation theory of the proper Lorentz group $\SOP$. The (complexified) Lorentz Lie algebra $\soc\cong\slc\oplus\slc$ is a direct sum of two Lie algebras of the complex special linear group $\SLC$~\cite{Ticciati}. This allows us to obtain all finite-dimensional representations, both Hermitian and non-Hermitian, of the Lorentz Lie algebra $\so$ from the finite-dimensional representations of $\slc$~\cite{Fuchs:1997jv}. We then exponentiate these to obtain all finite-dimensional representations of the proper Lorentz group $\SOP$. 

We use these pseudo-Hermitian representations to construct the dual quantum field operator for the simplest trivial representation $(0,0)$ of the spin-$0$ scalar field and a more complicated case of the smallest nontrivial representations $(\hf,0)$ and $(0,\hf)$, which represent spin-$\hf$ left- and right-handed spinors, respectively. Finally, we show how to construct pseudo-Hermitian representations and the dual quantum field for Dirac fermions, which are the direct sum $(0,\hf)\oplus (\hf,0)$ of right- and left-handed spinors.

\subsection{Irreducible representations of $\slc$}\label{sec:irreps}

The complex special linear group $\SLC$ is the group of complex $2\times 2$ matrices with a unit determinant:
\be\label{eq:definition_special_linear_group}
\SLC=\{A \in \GL \mid \text{det} (A)=1 \}\;.
\ee
It has the Lie algebra $\slc$, which is a complex vector space of traceless complex $2\times 2$ matrices:
\be\label{eq:slc_definition}
\slc=\{X \in \glc \mid \text{Tr}(X)=0 \}=\text{span}_{\cn}  \{ J_a\df \frac{\p{a}}{2}, \:a=1,2,3 \}\;.
\ee
The Lie algebra $\slc\cong \mathfrak{su}(2)_{\cn}$ is isomorphic to the complexified Lie algebra $\mathfrak{su}(2)_{\cn}$ of the special unitary group SU$(2)$. In simpler terms, $\slc$ is the vector space $\mathfrak{su}(2)$ spanned over complex numbers instead of real numbers. Thus, in the literature, it is common to write the Lorentz Lie algebra as a direct sum $\mathfrak{su}(2)\oplus\mathfrak{su}(2)$.

The generators of $\slc$ are $J_{a}=\frac{\p{a}}{2}$, where
$\p{a}$ are the Pauli matrices. The Lie brackets for this basis are
\be\label{eq:lie_brackets}
[J_{a},J_{b}]=i\ep_{abc}J_{c}\;.
\ee
The Lie algebra of $\slc$ possesses the structure of \comma\comma ladder operators'', which we obtain by defining a new basis:
\be\label{eq:ladder_basis}
\slc=\text{span}_{\cn}  \{\J{3}, \J{+}, \J{-}\} \text{\:\:\:with\:\:\:}
\J{\pm}\df \J{1}\pm i\J{2}\;.
\ee
Here, $\J{\pm}$ are the raising and lowering operators with the Lie brackets:
\be
[J_{3},J_{\pm}]=\pm J_{\pm}\;,\:\:\:[J_{+},J_{-}]=2J_{3}\;.
\ee

Now, let $ (\rho,\cv)$ be a representation of $\slc$ on some finite-dimensional vector space $\cv$:
\begin{equation}
\begin{array}{ll}
\rho: \slc\longmapsto \glc\:,\:\:\: & \rho(X): \cv \longmapsto \cv \\
\:\:\:\:\:\:\:\:\:\:\:\:\:X\:\:\: \longmapsto \rho(X) & \:\:\:\:\:\:\:\:\:\:\:\:\:\:v \longmapsto \rho(X)[v]\:.
\end{array}
\end{equation}
The dimension of a representation is defined as the dimension of the vector space $\text{dim}(\rho)=\text{dim}(\cv)$. It is also common to refer to the vector space $\cv$ as the \emph{representation space} of a representation $\rho$. In the context of \comma\comma standard" quantum field theory, we assume representations of $\slc$ to be Hermitian:
\be\label{eq:slc_Hermitian_rep_definition}
\rho(J_{a})^{\dag}=\rho(J_{a})\:,\:\:\:\forall a\in {1,2,3}\:.
\ee
Note that the representation $\rho(X)^{\dag}\neq \rho(X)$ on a general element $X\in\slc$ is not, in general, Hermitian, as $\slc$ is a complex vector space. Hence, the definition of a Hermitian representation is basis dependent. 

However, in general, the representations of $\slc$ need not be Hermitian. Given that $\re{J}{3}{}$ is finite dimensional and diagonalizable, we use the ladder operators (\ref{eq:ladder_basis}) to obtain all irreducible representations of $\slc$~\cite{Fuchs:1997jv}. An irreducible representation is a representation that cannot be broken down into smaller subset representations while preserving properties of the Lie algebra. 

All of the irreducible representations $ (\rho_{j},\cv_{j})$ of $\slc$ can be classified by an integer or half-integer number $j\in \{0,\frac{1}{2},1,...\}=\mathbb{N}/2$. The dimension of each representation is $\text{dim} (\rho_{j})=\text{dim} (\cv_{j})=2j+1$ and the representation space $\cv_{j}\subseteq \cn^{2j+1}$. 

Let us consider a representation $(\rho_{j},\cv_{j})$ and let $\hre{J}{3}{j}$ be non-Hermitian. As we assumed $\rho(J_{3})$ to be diagonalizable, we can use its eigenvectors to construct a biorthonormal basis \cite{Mostafazadeh:2001jk, Mostafazadeh:2008pw}, from which we obtain a Hermitian matrix $\pi:\cv_{j}\mapsto\cv_{j}$ such that $\hr{j}(J_{3})^{\dag}=\pi\hre{J}{3}{j}\pi^{-1}$ is $\pi$-pseudo-Hermitian. As representations preserve the Lie bracket:
\be
[\rho_{j}(J_{a}),\rho_{j}(J_{b})]=\rho_{j}([J_{a},J_{b}])=i \ep_{abc}\rho_{j}(J_{c})\:,
\ee
we have that the other generators are also $\pi$-pseudo-Hermitian:
\be\label{eq:definition_slc_pseudo_Hermitian}
\rho_{j}(J_{a})^{\dag}=\pi\hre{J}{a}{j}\pi^{-1}\:.
\ee
We call this representation a $\pi$-pseudo-Hermitian representation of $\slc$. Note that, for a general element $X\in\slc$ the representation $\rho_{j}(X)^{\dag}\neq \pi\rho_{j}(X)\pi^{-1}$ is not $\pi$-pseudo-Hermitian. Thus, the definition of a pseudo-Hermitian representation is basis dependent.

The smallest irreducible representations of $\slc$ are
\begin{enumerate}
    \item $j=0$,\:\:\: $\rho_{0}(J_{a})=0,\: \forall a \in \{1,2,3\}$, and $\cv_{0}\subseteq \cn$\: is the \emph{trivial} representation of $\slc$. It vanishes whether $(\rho_{0},\cv_{0})$ is Hermitian or non-Hermitian.
    \item $j=\hf$,\:\:\:$\rho^{\textsc{H}}_{\hf}(J_{a})=J_{a},\:\forall a\in \{1,2,3\}$, and $\cv_{\hf}\subseteq\cn^{2}$. Here, $(\rho^{\textsc{H}}_{\hf},\cv_{\hf})$ is the smallest nontrivial \emph{Hermitian} representation of $\slc$. 
    
    Any non-Hermitian representation $(\rho_{\hf},\cv_{\hf})$ will be related to the Hermitian representation via a similarity transformation:
\be\label{eq:similarity_transformation}
    \hre{J}{a}{\hf}=VJ_{a}V^{-1}\:.
    \ee
    We use a biorthonormal basis to construct a $2\times 2$ matrix $\pi:\cv_{\hf}\mapsto \cv_{\hf}$ such that $(\rho_{\hf},\cv_{\hf})$ is $\pi$-pseudo-Hermitian:
    \be
    \rho_{\hf}(J_{a})^{\dag}=\pi \rho_{\hf}(J_{a})\pi^{-1}\:.
    \ee
\end{enumerate}

\subsection{Irreducible representations of the proper Lorentz group}

Having classified all irreducible representations of $\slc$, we can obtain all irreducible representations of the proper Lorentz group $\SOP$. This follows from the fact that the complexified Lorentz Lie algebra $\soc\cong\slc\oplus\slc$ is a direct sum of two Lie algebras $\slc$ [or $\mathfrak{su}(2)_{\cn}$ depending on the literature].

The proper Lorentz group $\SOP$ is a Lie group of $4\times 4$ matrices:
\be
\SOP= \{\lt \in \text{GL} (4,\rn) \mid \ \lt^{\top}g \lt=g,\: \text{det} (\lt)=1 \text{\:\:and\:\:} \lt\indices{^0_0} \geq 1 \}\;.
\ee
Here, $g_{\mu\nu}=\text{diag}(1,-1,-1,-1)$ is the Minkowski metric.

The Lorentz Lie algebra $\so$ is a real vector space of traceless $4\times 4$ matrices:
\be
\so= \{M\in\mathfrak{gl} (4,\rn) \mid M=-g M^{\top}g,\: \text{Tr} (M)=0\}=\text{span}_{\rn} \{R_{a}, B_{a}, a=1,2,3  \}\;.
\ee
\noindent It has six generators:\ three rotation generators $R_{a}$ and three boost generators $B_{a}$, which form a basis of $\so$. The complexified Lorentz algebra is just the complex vector space:
\be
\soc=\text{span}_{\cn} \{R_{a}, B_{a}, a=1,2,3  \}\;.
\ee
\noindent Most literature on the representation theory does not distinguish between the Lie algebra and its complexification. This is because the majority of the results from the complexified Lie algebra can be directly applied to the original Lie algebra by taking the vector space to span over the real numbers instead of the complex numbers.

Let us define a new basis for the Lorentz Lie algebra $\soc$:
\be
T_{a}\df\hf (iR_{a}-B_{a})\:,\:\:\:K_{a}\df\hf (iR_{a}+B_{a})\:.
\ee
The Lie bracket for this basis is exactly that of $\slc\oplus\slc$: 
\be [T_{a},T_{b}]=i\ep_{abc}T_{c}\:,\:\:\: [K_{a},K_{b}]=i\ep_{abc}K_{c}\:,\:\:\: [T_{a},K_{b}]=0\;.
\ee
Hence, the Lorentz Lie algebra $\soc$ is isomorphic to the direct sum $\slc\oplus\slc$.

Now consider two irreducible representations $ (\rho_{j},\cv_{j})$ and $ (\rho_{k},\cv_{k})$ of $\slc$. We can construct a corresponding irreducible representation of $\slc\oplus\slc$:
\be
\ka_{jk}(X,Y)\df \rho_{j}(X)\otimes \id_{2k+1}+\id_{2j+1}\otimes\rho_{k}(Y)\text{\:\:\: for any two elements\:\:\:}X,Y\in\slc\:.
\ee
All irreducible representations of $\slc\oplus\slc$ are obtained from $(\ka_{jk},\cv_{jk})$, where the representation space of $\ka_{jk}$ is a tensor product $\cv_{jk}\df\cv_{j}\otimes\cv_{k}\subseteq \cn^{ (2j+1) (2k+1)}$. The dimension of this representation is $\text{dim}(\ka_{jk})=\text{dim}(\cv_{jk})=(2j+1)(2k+1)$. All irreducible representations of $\slc\oplus\slc$ are classified by two integer or half-integer numbers $(j,k)$, where $j,k\in\{0,\hf,1,\cdots\}=\mathbb{N}/2$.

The relationship between the elements of the complexified Lorentz Lie algebra $\soc$ and $\slc\oplus\slc$ is given explicitly through an isomorphism:
\be
\dr{T_{a}}{jk}\cong\ka_{jk}(J_{a},0)=\re{J}{a}{j}\otimes \id_{2k+1}\text{\:\:\:and\:\:\:}
\dr{K_{a}}{jk}\cong\ka_{jk}(0,J_{a})=\id_{2j+1}\otimes \re{J}{a}{k}\:.
\ee
This allows us to obtain all irreducible representations $(\ka_{jk},\cv_{jk})$ of the Lorentz Lie algebra $\so$ by going back to the basis of rotation and boost generators: 
\be\label{eq:reps_of_rots_and_boosts}
\begin{aligned}
&\dr{R_{a}}{jk}=-i [\re{J}{a}{j}\otimes \id_{2k+1}+\id_{2j+1}\otimes \re{J}{a}{k}]\:,\\
&\dr{B_{a}}{jk}=-\re{J}{a}{j}\otimes \id_{2k+1}+\id_{2j+1}\otimes \re{J}{a}{k}\:.
\end{aligned}
\ee
\noindent Thus, the $(j,k)$ representation for any element $M=\al^{a}R_{a}+\bt^{a}B_{a}\in\so$ with $\al^{a},\bt^{a}\in\rn$ of the Lorentz Lie algebra $\so$ is given by
\be\label{eq:reps_on_general_element}
\dr{M}{jk}=\re{-X^{\dag}}{}{j}\otimes\id_{2k+1}+\id_{2j+1}\otimes\re{X}{}{k}\:,
\ee
where we defined the coefficient $X^{a}\df\bt^{a}-i\al^{a}\in\cn$ so that $X=X^{a}J_{a}\in\slc$ is an element of the Lie algebra $\slc$. Hence, the Eq.~\eqref{eq:reps_on_general_element} explicitly maps an element of $\slc$ to an element of the Lorentz Lie algebra $\so$.

Let us consider two pseudo-Hermitian representations $(\rho_{j},\cv_{j})$ and $(\rho_{k},\cv_{k})$ of the Lie algebra $\slc$: 
\be
\rho_{j}(J_{a})^{\dag}=\pi_{j}\rho_{j}(J_{a})\pi_{j}^{-1}\text{\:\:\:and\:\:\:}\rho_{k}(J_{a})^{\dag}=\pi_{k}\rho_{k}(J_{a})\pi_{k}^{-1}\:,
\ee
where $\cv_{j}\subseteq \cn^{2j+1}$ and $\cv_{k}\subseteq \cn^{2k+1}$. Here, $\pi_{j}$ is some $(2j+1)\times (2j+1)$ Hermitian matrix and $\pi_{k}$ is a $(2k+1)\times(2k+1)$ Hermitian matrix, which can be found by constructing a biorthonormal basis and diagonalizing $\rho_{j}(J_{3})$ and $\rho_{k}(J_{3})$, respectively.

Now taking a Hermitian conjugate of this representation:
\be
\ka_{jk}(M)^{\dag}=-(\pi_{j}\otimes\pi_{k})\qty[\rho_{j}(X)\otimes\id_{2k+1}+\id_{2j+1}\otimes\rho_{k}(-X^{\dag})  ](\pi_{j}\otimes\pi_{k})^{-1}\:,
\ee
we find that it is related to the parity transformation of $M$:
\be
\ka_{jk}(\lt_{P}M\lt_{P})=\rho_{j}(X)\otimes \id_{2k+1}+\id_{2j+1}\otimes\rho_{k}(-X^{\dag})\:.
\ee
Here, $\lt_{P}=\text{diag}(1,-1,-1,-1)\in\SO$ is the parity matrix, which is an improper discrete Lorentz transformation. We used its action on the rotation and boost generators:
\be
\lt_{P}R_{a}\lt_{P}=R_{a}\text{\:\:\:and\:\:\:}\lt_{P}B_{a}\lt_{P}=-B_{a}\:.
\ee
However, the action of parity interchanges representation $(j,k)$ with $(k,j)$ and swaps the order of the tensor product, up to some similarity transformation:
\be
\ka_{jk}(\lt_{P}M\lt_{P})=S\ka_{kj}(M)S^{-1}\:,
\ee
where $S$ is some $(2j+1)(2k+1)\times (2j+1)(2k+1)$ similarity matrix. 

Thus, the $(j,k)$ representation of the Lorentz Lie algebra $\so$ will be anti-pseudo-Hermitian with respect to the interchange of $(j,k)$ and $(k,j)$:
\be
\ka_{jk}(M)^{\dag}=-\td{\pi}\ka_{kj}(M)\td{\pi}^{-1}\:,\text{\:\:\:where\:\:\:}\td{\pi}\df (\pi_{j}\otimes \pi_{k})S\:.
\ee

Let $ (D,\cv)$ be a representation of the proper Lorentz group $\SOP$ on some finite-dimensional vector space $\cv$:
\begin{equation}
\begin{array}{ll}
D:\ \SOP\longmapsto \text{GL}(\cv)\:,\:\:\:\:\:\:\: & D(\lt): \cv \longmapsto \cv \\
\:\:\:\:\:\:\:\:\:\:\:\:\:\:\:\:\:\:\lt\:\:\:\:\: \longmapsto D(\lt) & \:\:
\:\:\:\:\:\:\:\:\:\:\:\:v\: \longmapsto D(\lt)[v]\:.
\end{array}
\end{equation}
All finite-dimensional representations $(D_{jk},\cv_{jk})$ of the proper Lorentz group $\SOP$ can be obtained from the representations $(\ka_{jk},\cv_{jk})$ of the Lorentz Lie algebra $\so$ via the exponential map:
\be\label{eq:jk_rep_of_Lorentz_group}
D_{jk} (\lt)=e^{\dr{M}{jk}}=e^{\al^{a} \ka_{jk}(R_{a})+\bt^{a}\ka_{jk}(B_{a})}\;.
\ee
\noindent  We call this the $(j,k)$ representation of the proper Lorentz group $\SOP$ for some integer or half-integer numbers $j,k\in \{0,\hf,1,\cdots\}=\frac{\mathbb{N}}{2}$.

Given two pseudo-Hermitian representations of $\slc$, the corresponding $(j,k)$ representation of the proper Lorentz group $\SOP$ will be pseudounitary with respect to the interchange of $(j,k)$ and $(k,j)$: 
\be
D_{jk}(\lt)^{\dag}=e^{\ka_{jk}(M)^{\dag}}=\td{\pi}e^{-\ka_{kj}(M)}\td{\pi}=\td{\pi}D_{kj}(\lt)^{-1}\td{\pi}^{-1}\:.
\ee
Hence, all finite-dimensional pseudounitary representations of the proper Lorentz group $\SOP$ can be obtained from finite-dimensional pseudo-Hermitian representations of the Lie algebra $\slc$. 

In the following subsections we look for pseudounitary representations of $\SOP$ for the trivial representation $(0,0)$ of scalar fields and the smallest nontrivial representations $(0,\hf)$ and $(\hf,0)$ of right- and left-handed spinors, as well as $(0,\hf)\oplus(\hf,0)$ representation of fermions.

\subsection{Scalar fields}\label{sec:scalar_irreps}

Scalar fields transform in the trivial $(0,0)$ representation of the proper Lorentz group $\SOP$. The representation space $\cv_{00}=\cv_{0}\otimes\cv_{0}\subseteq \cn$ is just a subset of the complex numbers. The $j=0$ representation of the Lie algebra $\slc$ vanishes $\rho_{0}(J_{a})=0$ for all generators of $\slc$~\cite{Ticciati}, regardless of whether it is Hermitian or non-Hermitian. Thus, from Eq.~\eqref{eq:reps_of_rots_and_boosts}, we see that the trivial representation of rotation and boost generators $\ka_{00}(R_{a})=\ka_{00}(B_{a})=0$ also vanishes. Thus, the corresponding $(0,0)$ representation of the proper Lorentz group $\SOP$ is found via the exponential map in Eq.~\eqref{eq:jk_rep_of_Lorentz_group} to be unity:
\be
D_{00}(\lt)=e^{\al^{a}\ka_{00}(R_{a})+\bt^{a}\ka_{00}(B_{a})}=e^{0}=1,\:\:\:\forall \lt\in\SOP\;.
\ee
\noindent Hence, $D_{00} (\lt)^{\dag}=\pi D_{00} (\lt)^{-1}\pi^{-1}=\pi \pi^{-1}=1$ is $\pi$-pseudo-unitary for any real number $\pi\in\rn$ (as $\pi$ needs to be Hermitian). 

We use Eq.~\eqref{eq:main_result_1} to define the dual scalar field operator:
\be
\thph{\dag}{}{x}=\heta^{-1}\hphd (x^{\eta})\heta\: \pi\;,\text{\:\:\:with\:\:\:}\pi\in \rn \;.
\ee

For a single-component scalar field, $\pi\in\rn$ can be any real number. However, this is not true if we consider multicomponent scalar/pseudoscalar fields. In the case of an $n$-component scalar field $\hph^{a} (x)$, $a\in\{1,2,\cdots,n\}$, the dual $n$-component field is defined via
\begin{equation}\label{eq:dual_field_for_n_comp_scalars}
\left.\begin{array}{c}
\thph{\dag}{1}{x}=\heta^{-1}\hph^{\dag }_{1} (x^{\eta})\heta \: \pi_{1} \\
\thph{\dag}{2}{x}=\heta^{-1}\hph^{\dag }_{2} (x^{\eta})\heta \: \pi_{2} \\
\vdots \\
\thph{\dag}{n}{x}=\heta^{-1}\hph^{\dag }_{n} (x^{\eta})\heta \: \pi_{n}
\end{array}\right\} \Rightarrow \thph{\dag}{}{x}=\heta^{-1}\hph (x^{\eta})\heta\:\Pi\;.
\end{equation}
\noindent Here, $\hph (x)=\mqty(\hph^{1} (x)\\ \vdots\\ \hph^{n} (x))$ is an $n$-component scalar/pseudoscalar field and $\Pi=\mqty(\pi_{1}&0&\cdots&0\\
0&\pi_{2}& &\\
\vdots & & \ddots & \\
0 & & &\pi_{n})$ is a diagonal matrix of real numbers $\pi_{1},\pi_{2},\cdots,\pi_{n}\in\rn$.

Returning to the construction of pseudo-Hermitian quantum field theories, we require the Lagrangian operator $\hlag (x)$ to be
\begin{enumerate}
    \item Invariant under proper Poincar\'{e} transformations:
    
    I$\SOP$:\: $\hlag (x)\longmapsto \hu^{-1} (\ep,\lt)\hlag (x)\hu (\ep,\lt)=\hlag (\lt^{-1}(x-\ep))$\:,
    \item 
    $\heta$-pseudo-Hermitian:\: $\hlag (x^{\eta})^{\dag}=\heta\hlag (x)\heta^{-1}$.
\end{enumerate}
In the case of noninteracting theories, the kinetic and diagonal mass terms are of the form $\thph{\dag}{}{x}\hph (x)$, which are both Poincar\'{e} invariant and $\heta$-pseudo-Hermitian. More generally, however, we can include a mass mixing. In this case, there will be a nondiagonal mass term of the from $\thph{\dag}{}{x}B\hph (x)$, where $B$ is some $n\times n$ nondiagonal matrix. Under the action of the proper Poincar\'e group, we have
\be
\text{I}\SOP: \thph{\dag}{}{x}B\hph (x) \mapsto \thph{\dag}{}{\lt^{-1}(x-\ep)} [D^{-1} (\lt)BD (\lt)]\hph (\lt^{-1}(x-\ep))\;.
\ee
For $n$-component scalars/pseudoscalars, the $n$-dimensional proper Lorentz transformation is just the $n\times n$ identity matrix, i.e.,
\be
D (\lt)=\mqty(D_{00}(\lt)&0&\cdots&0\\
0&D_{00}(\lt)& &\\
\vdots & & \ddots & \\
0 & & &D_{00}(\lt))=\id_{n}\:.
\ee
\noindent Hence, the mass mixing terms are Poincar\'e invariant. However, the mixing term is $\heta$-pseudo-Hermitian:
\be
(\thph{\dag}{}{x^{\eta}}B\hph (x^{\eta}))^{\dag}=\heta \qty[\thph{\dag}{}{x}\Pi^{-1}B^{\dag}\Pi \hph (x)] \heta^{-1}=\heta \qty[\thph{\dag}{}{x}B\hph (x)]\heta^{-1}\;,
\ee
if and only if $B^{\dag}=\Pi B \Pi^{-1}$ is $\Pi$-pseudo-Hermitian. This places restrictions on the matrices $B$ and $\Pi$. 

In Sec.~\ref{sec:PT_example}, we study an example of this type of Lagrangian. We consider a 2-component scalar field with mass mixing term of the form $\hlag (x)=\thph{\dag 1}{}{x}\hph^{2} (x)-\thph{\dag 2}{}{x}\hph^{1} (x)$. In matrix form this term is
\begin{equation}  
B=\mqty(0&1\\-1&0)\;.
\end{equation}
However, the matrix $B$ is $\Pi$-pseudo-Hermitian if and only if\: $\pi_{2}=-\pi_{1}$, which restricts $\Pi$ to the form:
\begin{equation}
\Pi\propto \mqty(1&0\\0&-1)\;.
\end{equation}
\noindent In the example, this is a parity matrix that reflects the intrinsic parities of a scalar component $\hph^{1}$ and a pseudoscalar component $\hph^{2}$ in the 2-component scalar field $\hph=\mqty(\hph^{1}\\ \hph^{2})$.

\subsection{Two-component spinors}\label{sec:spinors}

Unlike for the case of scalar fields, the representations $(\hf,0)$ and $(0,\hf)$, which correspond to left- and right-handed spinors, are nontrivial. The dual left- and right-handed field operators are obtained via Eq.~\eqref{eq:main_result_1}. However, introduction of the dual field alone is insufficient to formulate a Lorentz-invariant spinor Lagrangian. This is due to the dependence of the kinetic terms on the Pauli matrices, which do not transform in the same representation of $\SOP$ as the quantum field operators. Our goal is to write down the spinor Lagrangian which transforms in the same representation of the proper Poincar\'e group I$\SOP$. Having the spinor Lagrangian in hand, we obtain a pseudo-Hermitian representation for Dirac fermions $(0,\hf)\oplus (\hf,0)$. However, a comprehensive study of a specific example of a non-Hermitian fermionic quantum field theory is beyond the scope of this work.

The smallest nontrivial irreducible representations of the Lorentz Lie algebra $\so$ are $(\hf,0)$, acting on the left-handed spinor $\hps_{L}$, and $(0,\hf)$, acting the right-handed spinor $\hps_{R}$:
 \be
\begin{aligned}
\hu^{-1}(\lt)\hps_{L}(x)\hu(\lt)&= D_{\hf 0}(\lt)\hps_{L}(\lt^{-1}x)\:,\\
\hu^{-1}(\lt)\hps_{R}(x)\hu(\lt)&= D_{ 0\hf}(\lt)\hps_{R}(\lt^{-1}x)\: .
\end{aligned}
\ee
We obtain the $2\times 2 $ matrix representations $(D_{\hf 0},\cn^{2})$ and $(D_{0 \hf},\cn^{2})$ of the proper Lorentz group $\SOP$ via the exponential map in Eq.~\eqref{eq:jk_rep_of_Lorentz_group}:
\be
D_{\hf 0}(\lt)=e^{\ka_{\hf 0}(M)}\text{\:\:\:and\:\:\:}D_{0\hf}(\lt)=e^{\ka_{0\hf}(M)}\:,
\ee
where $M=\al^{a}R_{a}+\bt^{a}B_{a}\in\so, \al^{a},\bt^{a}\in\rn$ is a general element in the Lorentz Lie algebra $\so$. The $2\times 2$ matrix representations $(\ka_{\hf 0},\cn^{2})$ and $(\ka_{0\hf},\cn^{2})$ of the Lorentz Lie algebra $\so$ are related to the $2\times 2$ matrix representation $(\rho_{\hf},\cn^{2})$ of $\slc$ via Eq.~\eqref{eq:reps_on_general_element}:
\be\label{eq:spinor_reps_on_general_element_M}
\ka_{0\hf}(M)=\rho_{\hf}(X) \text{\:\:\:and\:\:\:} \ka_{\hf 0}(M)=\rho_{\hf}(-X^{\dag})\:.
\ee
We can map any element in the Lorentz Lie algebra $M\in\so$ to an element $X\in\slc$.

If the representation $\big(\rho_{\hf},\cn^{2})$ is non-Hermitian, we have seen that we can use a biorthonormal basis ~\cite{Mostafazadeh:2001jk, Mostafazadeh:2008pw} to construct a $2\times 2$ Hermitian matrix $\pi:\cn^{2}\mapsto\cn^{2}$ such that the representation is $\pi$-pseudo-Hermitian:
\be
\rho_{\hf} (J_{a})^{\dag}=\pi\hre{J}{a}{\hf}\pi^{-1}\;.
\ee
This leads to Lorentz Lie algebra representations being anti-$\pi$-pseudo-Hermitian under an interchange of left-handed $(\hf,0)$ and right-handed $(0,\hf)$ representations:
\be
\dr{M}{\hf 0}^{\dag}=-\pi\dr{M}{0 \hf}\pi^{-1}
\text{\:\:\:and\:\:\:}
\dr{M}{0 \hf}^{\dag}=-\pi\dr{M}{\hf 0}\pi^{-1}\;.
\ee
\noindent Hence, the representations of the proper Lorentz group are $\pi$-pseudo-unitary under the interchange of left-handed and right-handed representations:
\be
\lr{\hf 0}^{\dag}=\pi \lr{0\hf}^{-1}\pi^{-1}\;,\text{\:\:\:\:\:}
\lr{0 \hf}^{\dag}=\pi \lr{\hf 0}^{-1}\pi^{-1}\;.
\ee

As shown in Sec.~\ref{sec:Fock_space_reps} we can write down the dual field operator, provided the finite-dimensional representations are pseudounitary. Thus, we define the dual left- and right handed quantum field operators using Eq.~\eqref{eq:main_result_1}:
\be
\hat{\td{\psi}}^{\dag}_{L} (x)=\heta^{-1}\hpsd_{L} (x^{\eta})\heta\:\pi \text{\:\:\:and\:\:\:} \hat{\td{\psi}}^{\dag}_{R} (x)=\heta^{-1}\hpsd_{R} (x^{\eta})\heta\:\pi\;,
\ee
where $\heta: x\mapsto x^{\eta}$ is the coordinate transformation of $x\in\rn^{1,3}$ under $\heta$.

Indeed, we can check that, under the proper Lorentz group $\SOP$, the left-handed dual field $\hat{\td{\psi}}_{L}$ transforms under the dual representation of the right-handed spinor $\hps_{R}$ and the right-handed dual field $\hat{\td{\psi}}_{R}$ transforms under the dual representation of the left-handed spinor $\hps_{L}$:
\be
\begin{aligned} 
\hu^{-1}(\lt)\hat{\td{\psi}}^{\dag}_{L}(x)\hu(\lt)&=\hat{\td{\psi}}^{\dag}_{L}(\lt^{-1} x)D_{0\hf}(\lt)^{-1}\:,\\
\hu^{-1}(\lt)\hat{\td{\psi}}^{\dag}_{R}(x)\hu(\lt)&=
\hat{\td{\psi}}^{\dag}_{R}(\lt^{-1} x)D_{\hf 0}(\lt)^{-1}\:.
\end{aligned}
\ee
Hence, the typical mass terms in the non-Hermitian spinor Lagrangian
\be\label{eq:spinor_mass_terms}
\hat{\td{\psi}}_{L}^{\dag}(x)\hps_{R}(x)\text{\:\:\:and\:\:\:}\hat{\td{\psi}}_{R}^{\dag}(x)\hps_{L}(x)
\ee
\noindent are both $\heta$-pseudo-Hermitian and invariant under proper Poincar\'{e} transformations. Moreover, in the Hermitian limit, where $\heta=\hat{\id}$ and $\pi=\id_{2}$, we recover the mass terms of the Hermitian spinor Lagrangian:
\be
\hpsd_{L}(x)\hps_{R}(x)\text{\:\:\:and\:\:\:}\hpsd_{R}(x)\hps_{L}(x)\:.
\ee

However, even with the dual spinor fields, the kinetic terms will not be Lorentz invariant. This is because the kinetic terms involve Minkowski 4-vectors in the Hermitian representation of $\slc$. Consider the kinetic terms in the Hermitian spinor Lagrangian~\cite{Ticciati}
\be
\hpsd_{L}(x)[\partial]\hps_{L}(x)\text{\:\:\:and\:\:\:} \hpsd_{R}(x)[\partial]_{P}\hps_{R}(x)\:.
\ee
Here, $[\partial]$ is a map from Minkowski 4-vectors to $2\times 2$ matrices defined as (for more details see Sec.\:7 in Ref.~\cite{Ticciati}):
\be\label{eq:map_SO_to_SLC}
\begin{aligned}
    [\:]:\: &\rn^{1,3} \longmapsto \text{Mat}_{2}(\rn^{1,3})\:,\\
    &\partial \:\:\:\:\:\longmapsto [\partial]\df\p{\mu}\partial^{\mu}\:,\\
    &\partial_{P} \:\:\:\longmapsto [\partial]_{P}\df\bar{\sigma}_{\mu}\partial^{\mu}\:.
\end{aligned}
\ee
\noindent We have defined $\partial^{\mu}_{P}=(\partial^{0},-\partial^{i})$ to be the parity transformation of the 4-vector $\partial^{\mu}$, and the four-Pauli matrices are $\sigma_{\mu}=(\id_{2},\sigma_{i})$ and $\bar{\sigma}_{\mu}=(\id_{2},-\sigma_{i})$.

The map $[\:]$ gives an explicit relationship between the proper Lorentz group $\SOP$ and the complex special linear group $\SLC$:
\be\label{eq:explicit_map_to_SLC}
\begin{aligned}
    \SOP:\: &\text{Mat}_{2}(\rn^{1,3})&&\longmapsto \text{Mat}_{2}(\rn^{1,3})\:,\\\
    &[\partial]
&&\longmapsto [\lt\cdot\partial]=A[\partial]A^{\dag}\:,\\
    &[\partial]_{P}&&\longmapsto [\lt\cdot\partial]_{P}=A^{-1 \dag}[\partial]_{P}A^{-1}\;.
\end{aligned}
\ee
Here, $A\in\SLC$ is a $2\times 2$ matrix in the complex special linear group, see Eq.~\eqref{eq:definition_special_linear_group}. Indeed, we can write any element of $\SLC$ in terms of its Lie algebra generators $X\in\slc$:
\be
A=e^{X}=e^{\rho^{\textsc{H}}_{\hf}(X)}=e^{\ka^{\textsc{H}}_{0\hf}(M)}=D^{\textsc{H}}_{0\hf}(\lt)\:,
\ee
where we have used Eq.~\eqref{eq:spinor_reps_on_general_element_M}, which relates the $j=\hf$ representation of $\slc$ to the $(0,\hf)$ representation of $\so$. We also note that $(\rho^{\textsc{H}}_{\hf},\cn^{2})$ is the Hermitian $j=\hf$ representation of $\slc$, meaning that $\rho^{\textsc{H}}_{\hf}(J_{a})^{\dag}=\rho^{\textsc{H}}_{\hf}(J_{a})=J_{a}$. Thus, we also denote $(D^{\textsc{H}}_{0\hf},\cn^{n})$ to be the unitary representation of $\SOP$ with respect to the interchange of $(0,\hf)$ and $(\hf,0)$, such that $D^{\textsc{H}}_{0\hf}(\lt)^{\dag}=D^{\textsc{H}}_{\hf 0}(\lt)^{-1}$ [note that $A\in\SLC$ is not unitary].
Hence, the derivative matrices $[\partial]$ and $[\partial]_{P}$ transform in the unitary representation of the proper Lorentz group, while the spinor fields transform in the pseudounitary representation. Due to this discrepancy, the kinetic terms will not be Lorentz invariant:
\begin{align}
\SOP:\:&\hat{\td{\psi}}_{L}^{\dag}\bqty{\partial}\hps_{L} \mapsto \hat{\td{\psi}}_{L}^{\dag}\underbrace{\lr{0\hf}^{-1} \plr{0 \hf}}_{\neq \id_{2}}\bqty{\partial}\underbrace{\plr{\hf 0}^{-1}\lr{\hf 0}}_{\neq \id_{2}}\hat{\psi}_{L}\:,\\
\:&\hat{\td{\psi}}_{R}^{\dag}\bqty{\partial}_{P}\hps_{R} \mapsto \hat{\td{\psi}}_{R}^{\dag}\lr{\hf 0}^{-1} \plr{ \hf 0}\bqty{\partial}_{P}\plr{ 0 \hf}^{-1}\lr{ 0\hf}\hat{\psi}_{R}\;.
\end{align}

However, the $\pi$-pseudo-Hermitian representation $(\rho_{\hf},\cn^{2})$ is related to the Hermitian representation $(\rho^{\textsc{H}}_{\hf},\cn^{2})$ by a similarity transformation (\ref{eq:similarity_transformation}):
\be
\re{X}{}{\hf}=V\rho^{\textsc{H}}_{\hf}(X)V^{-1}=VXV^{-1}\;.
\ee
Thus, the $\pi$-pseudo-unitary representation of the proper Lorentz group $\SOP$ is related to the unitary one via a similarity transformation:
\be
\lr{0\hf}=V\plr{0\hf}V^{-1}\text{\:\:\:and\:\:\:}D_{ \frac{1}{2} 0 }(\Lambda)= ( V^{\dag}\pi  )^{-1} D_{ \frac{1}{2}  0 }^{H}(\Lambda) (V^{\dag}\pi) .
\ee

Using this property, we define a new map from Minkowski 4-vectors to $2\times 2$ matrices:
\begin{align}\label{eq:new_derivative_map_curly_braket}
    \{\:\}:\:&\rn^{1,3} \longmapsto \text{Mat}_{2}(\rn^{1,3})\:,\\
    &\partial \:\:\:\:\:\longmapsto \{\partial\}\df V[\partial]V^{\dag}\pi=V\sigma_{\mu}V^{\dag}\pi\:\partial^{\mu}\:,\\
    &\partial_{P} \:\:\:\longmapsto \{\partial\}_{P}\df\pi^{-1}V^{\dag -1}[\partial]_{P}V^{-1}=\pi^{-1}V^{\dag -1}\bar{\sigma}_{\mu}V^{-1}\:\partial^{\mu}\:.
\end{align}
We can check that this map indeed transforms in the $\pi$-pseudo-unitary representation of the proper Lorentz group:
\be
\begin{aligned}
    \SOP:\: &\text{Mat}_{2}(\rn^{1,3})&&\longmapsto \text{Mat}_{2}(\rn^{1,3})\:,\\\
    &\{\partial\}
&&\longmapsto \{\lt\cdot\partial\}=D_{0\hf}(\lt)\{\partial\}D_{\hf 0}(\lt)^{-1}\:,\\
    &\{\partial\}_{P}&&\longmapsto \{\lt\cdot\partial\}_{P}=D_{\hf 0}(\lt)\{\partial\}_{P}D_{0\hf}(\lt)^{-1}\;.
\end{aligned}
\ee
Hence, the kinetic terms in a non-Hermitian spinor Lagrangian, which are both $\heta$-pseudo-Hermitian and Lorentz invariant, will be of the form
\be\label{eq:kinetic_spinor_terms}
\hat{\td{\psi}}_{L}^{\dag}(x) \{\partial\}\hps_{L}(x)\text{\:\:\:and\:\:\:}
\hat{\td{\psi}}_{R}^{\dag}(x)\{\partial\}_{P}\hps_{R}(x)\;.
\ee
Thus, combining the mass terms in Eq.~\eqref{eq:spinor_mass_terms} with the kinetic terms in Eq.~\eqref{eq:kinetic_spinor_terms} we can write down the noninteracting part of the non-Hermitian spinor Lagrangian as
\be\label{eq:spinor_Lagrangian}
\hlag(x)=\hat{\td{\psi}}_{L}^{\dag}(x) \{\partial\}\hps_{L}(x)+\hat{\td{\psi}}_{R}^{\dag}(x)\{\partial\}_{P}\hps_{R}(x)-m\hat{\td{\psi}}_{L}^{\dag}(x)\hps_{R}(x)-m\hat{\td{\psi}}_{R}^{\dag}(x)\hps_{L}(x)\:.
\ee
This Lagrangian is both $\heta$-pseudo-Hermitian and invariant under proper Poincar\'e transformations I$\SOP$, as required. Moreover, in the Hermitian limit, it reduces to the free Hermitian spinor Lagrangian.

\subsection{Four-component fermions}\label{4-comp-fermions}

Having found the pseudo-Hermitian representations of right- and left-handed spinors, we are able to obtain the corresponding pseudo-Hermitian representation of Dirac fermions. The Dirac fermion is a $4$-component field, which transforms under the direct sum $(0, \hf)\oplus (\hf, 0)$ of right-handed and left-handed spinor representations of the proper Lorentz group~\cite{Ticciati}.

We begin by defining a new representation on the Lorentz Lie algebra $\so$ by taking a direct sum of the right- and the left-handed spinor representations $s\df\kappa_{ 0\hf}\oplus\kappa_{\hf 0}$ over the direct sum of their representation spaces $\cn^{2}\oplus\cn^{2}\cong\cn^{4}$:
\be
\begin{aligned}
s(M) : \:&\cn^{2}\oplus\cn^{2}\longmapsto \cn^{4}\:,\\
& \hps_{R}\oplus\hps_{L} \longmapsto s(M)[\hps_{R}\oplus\hps_{L}]=(\ka_{ 0\hf}\oplus\ka_{\hf 0})(M)[\hps_{R}\oplus\hps_{L}]\:.
\end{aligned}
\ee
As $s(M)$ is a direct sum, it can be written as $4\times 4$ block-diagonal matrix acting on a $4$-component quantum field composed of $2$-component spinors:
\be\label{eq:s_rep_definition}
s(M)[\hps_{R}\oplus\hps_{L}]\cong\mqty(\dr{M}{ 0\hf}&0\\0&\dr{M}{\hf 0})\mqty(\hps_{R}\\ \hps_{L})=\mqty(\hre{X}{}{\hf }&0\\0&\hre{-X^{\dag}}{}{\hf})\mqty(\hps_{R} \\ \hps_{L})\:.
\ee
Here, we used Eq.~\eqref{eq:spinor_reps_on_general_element_M}, which relates the $j=\hf$ representation of $\slc$ to the $(0,\hf)$ and $(\hf,0)$ representations of the Lorentz algebra $\so$.

A more familiar basis is defined by acting on the rotation and boost generators:
\be
S_{ab}\df i \ep_{abc}s (R_{c})\text{\:\:\:and\:\:\:}
S_{0a}\df -is (B_{a})\;.
\ee
It obeys the Lorentz Lie bracket in Eq.~\eqref{eq:Poincare_Lie_Braket} and can be written as the commutator of the gamma matrices:
\be
S_{\mu \nu}=\frac{i}{4}\bqty{\td{\gm}_{\mu},\td{\gm}_{\nu}}\:.
\ee
However, here, the gamma matrices $\td{\gamma}_{a} \:(a=1,2,3)$ are not in the Hermitian representation of $\slc$. Instead, they are in the $(\rho_{\hf},\cn^{2})$ $\pi$-pseudo-Hermitian representation of $\slc$. This is due to the pseudo-Hermitian nature of the representation $(s,\cn^{4})$ defined in Eq.~\eqref{eq:s_rep_definition}. Hence, in the Weyl (chiral) basis, the gamma matrices are given by
\be
\td{\gm}_{\mu}=\mqty(0 & \re{\sigma}{\mu}{\hf} \\ \re{\bar{\sigma}}{\mu}{\hf}&0)\;.
\ee
Note that $\rho_{\hf}(\sigma_{0})=\rho_{\hf}(\id_{2})\notin\slc$ is not an element of the Lie algebra $\slc$. Instead, we have a Lie bracket
\be
[\rho_{\hf}(\sigma_{\mu}),\rho_{\hf}(\sigma_{\nu})]=2i\:\delta\indices{_\mu_a}\delta\indices{_\nu_b}\:\ep_{abc}\:\rho_{\hf}(\sigma_{c})\:.
\ee
The $\pi$-pseudo-Hermitian representation will be related to the Hermitian representation via a similarity transformation:
\be
\rho_{\hf}(\sigma_{\mu})=V\sigma_{\mu}V^{-1} \implies \rho_{\hf}(\sigma_{0})=\sigma_{0}=\id_{2}\:.
\ee
Thus, the gamma matrix $\td{\gm}_{0}$ is the same for Hermitian and $\pi$-pseudo-Hermitian representation. We can also check that the gamma matrices $\td{\gm}_{\mu}$ obey the Clifford algebra
\be
\{\td{\gm}_{\mu},\td{\gm}_{\nu}\}=2g_{\mu\nu}\id_{4}\:,
\ee
where $g_{\mu\nu}=\text{diag}(1,-1,-1,-1)$ is the Minkowski metric.

The Poincar\'e invariant, pseudo-Hermitian spinor Lagrangian in Eq.~\eqref{eq:spinor_Lagrangian} can be written in the 4-component Weyl (chiral) basis as
\be
\hlag =\mqty(\hat{\td{\psi}}^{\dag}_{R}&\hat{\td{\psi}}^{\dag}_{L})\mqty(\{\partial\}_{P}&0\\ 0 &\{\partial\})\mqty(\hps_{R}\\ \hps_{L})-m\mqty(\hat{\td{\psi}}^{\dag}_{R}&\hat{\td{\psi}}^{\dag}_{L})\mqty(0&\id_{2}\\ \id_{2} &0)\mqty(\hps_{R}\\ \hps_{L})\:.
\ee
We define the $4$-component Dirac fermion and its dual as
\be
\hps(x)\df \mqty(\hps_{R}(x)\\ \hps_{L}(x))\:,\:\hat{\td{\psi}}^{\dag}(x)\df \mqty(\hat{\td{\psi}}^{\dag}_{R}(x)&\hat{\td{\psi}}^{\dag}_{L}(x))\text{\:\:\:and\:\:\:}\hat{\bar{\td{\psi}}}(x)\df\hat{\td{\psi}}^{\dag}(x)\td{\gamma}_{0}\:.
\ee
If we expand the partial derivative matrices in terms of their definition in Eq.~\eqref{eq:new_derivative_map_curly_braket}, then we can write the Lagrangian in terms of gamma matrices:
\be
\hlag (x)=\hat{\bar{\td{\psi}}}(x)\mqty((V V^{\dag}\pi)^{-1} & 0\\ 0 & V V^{\dag}\pi)\td{\gamma}_{\mu}\partial^{\mu}\hps (x) - m\hat{\bar{\td{\psi}}}(x)\hps (x)\:.
\ee
As we see, while the mass term is of the usual form of the Dirac Lagrangian, the kinetic term has picked up a matrix $VV^{\dag}\pi$. Here, $\pi$ is the $2\times 2$ matrix constructed using a biorthonormal basis for $\slc$ and $V$ is the matrix diagonalizing the representation of $\rho_{\hf}(J_{3})$. Hence, if the eigenstates of $\rho_{\hf}(J_{3})$ have a positive norm with respect to the inner product $\braket{\cdot}{\pi \cdot}$, then $VV^{\dag}\pi=\id_{2}$ is just the identity. On the other hand, if the eigenstates have indefinite norm with respect to $\braket{\cdot}{\pi \cdot}$, then $VV^{\dag}\pi=C$ is the $C$ matrix, i.e., the discrete symmetry of $\rho_{\hf}(J_{3})$, meaning $[\rho_{\hf}(J_{3}),C]=0$, which we find from the biorthonormal basis for our system~\cite{Mostafazadeh:2001jk, Mostafazadeh:2008pw}. We call this matrix $C$ due to convention across the literature of non-Hermitian quantum mechanics~\cite{Bender:2002vv}; however, it has nothing to do with charge conjugation.

\section{Example: $PT$-Symmetric Scalar Field Theory}\label{sec:PT_example}

Before concluding this work, we consider a concrete example of a $PT$-symmetric scalar field theory. The theory is composed of two complex scalar fields with a non-Hermitian mass mixing matrix. This archetypal  model, introduced in Ref.~\cite{Alexandre:2017foi}, has been considered in a number of existing works (see, e.g., Refs.~\cite{Alexandre:2018uol, Alexandre:2020gah}). In this section, we show how it can be formulated consistently based on our preceding discussions.

\subsection{Naive Lagrangian}\label{sec:naive_lagrangian}

Following the strategy of appending non-Hermitian terms to an otherwise Hermitian Lagrangian, it is tempting to write the Lagrangian density, as was done in Ref.~\cite{Alexandre:2017foi}:
\be
\lag (x)=\partial^{\mu}\Phi^{\dag}(x)\partial_{\mu}\Phi (x)-\Phi^{\dag}(x)M^{2}\Phi (x) \;.
\ee
Here, $\Phi=\mqty(\Phi_{1}\\ \Phi_{2})$ is a 2-component complex scalar field composed of a scalar component $\Phi^{1}$ and a pseudoscalar component $\Phi^{2}$. The Lagrangian's non-Hermiticity arises from the presence of a non-Hermitian mass matrix:
\begin{equation}\label{eq:mass_mixing_matrix}
M^{2}=\mqty(m^{2}_{1}&\mu^{2}\\-\mu^{2}&m_{2}^{2})\neq M^{2^\dag}\;.
\end{equation}

Naively, we might assume the standard parity $P$ and time-reversal $T$ transformation properties of scalar and pseudoscalar fields as follows:
\begin{align}\label{eq:naive_PT_transformations_classical}
&\left.\begin{array}{l}
P:\:\Phi^{1}(x) \longmapsto \Phi^{1^P}(x^{P})=+\Phi^{1}(x) \\
\:\:\:\:\:\:\:\:\Phi^{2}(x) \longmapsto \Phi^{2^P}(x^{P})=-\Phi^{2}(x)
\end{array}\right\} \Rightarrow \Phi (x) \longmapsto \Phi^{P}(x^{P})=P\Phi (x)\;,\\
&\left.\begin{array}{l}
T:\:\Phi^{1}(x) \longmapsto \Phi^{1^T}(x^{T})=+\Phi^{1^*}(x) \\
\:\:\:\:\:\:\:\:\Phi^{2}(x) \longmapsto \Phi^{2^T}(x^{T})=+\Phi^{2^*}(x)
\end{array}\right\} \Rightarrow \Phi (x) \longmapsto \Phi^{T}(x^{T})=\Phi^{*} (x)\;.
\end{align}
Here, $P$ is the \comma`parity'' matrix:
\begin{equation}\label{eq:parity_matrix}
    P=\mqty(1 &0\\0&-1)\;,
\end{equation}
reflecting the intrinsic parity $+1$ of the scalar and $-1$ of the pseudoscalar field. Furthermore, we observe that the mass matrix is $P$-pseudo-Hermitian, i.e., $M^{2^\dag}=P M^{2}P^{-1}$, with respect to the parity matrix $P$. Indeed, the classical Lagrangian is $PT$ symmetric under the standard parity and time-reversal transformations in Eq.~\eqref{eq:naive_PT_transformations_classical}:
\be
PT: \lag (x) \longmapsto \lag^{PT}(x^{PT})=\lag (x)\;.
\ee

Let us now consider the corresponding quantum Lagrangian operator $\hlag:\fs\mapsto\fs$, acting on the Fock space $\fs$:
\be\label{eq:naive_quantum)lagrangian}
\hlag(x)=\der{}{\mu}{\hphd}(x)\der{\mu}{}{\hph}(x)-\hphd(x)M^{2}\hph (x)\;.
\ee
Here, $\hph=\mqty(\hph_{1}\\ \hph_{2})$ is the 2-component quantum field operator composed of a scalar field operator $\hph^{1}$ and a pseudoscalar field operator $\hph^{2}$. The corresponding standard scalar and pseudoscalar transformations under parity and time-reversal are
\begin{align}\label{eq:naive_parity_and_time_reversal_quantum}
&\left.\begin{array}{l}
P:\:\hpr \hph^{1}(x)\hpr^{-1}=+\hph^{1}(x^{P}) \\
\:\:\:\:\:\:\:\:\hpr\hph^{2}(x)\hpr^{-1}=-\hph^{2}(x^{P})
\end{array}\right\} \Rightarrow  \hpr\hph(x)\hpr^{-1}=P\hph (x^{P})\;,\\
&\left.\begin{array}{l}
T:\:\htr\hph^{1}(x)\htr^{-1}=+\hph^{1}(x^{T}) \\
\:\:\:\:\:\:\:\:\htr\hph^{2}(x)\htr^{-1}=+\hph^{2}(x^{T})
\end{array}\right\} \Rightarrow \htr\hph(x)\htr^{-1}=\hph (x^{T})\;,
\end{align}
where $\hpr:\fs\mapsto\fs$ is the parity operator and $\htr:\fs\mapsto\fs$ is the time-reversal operator, both acting on Fock space $\fs$. Indeed, the quantum Lagrangian is $\hpr$-pseudo-Hermitian under the standard parity transformation in Eq.~\eqref{eq:naive_parity_and_time_reversal_quantum}:
\be
\hlag (x^{P}) ^{\dag}=\hpr \hlag (x) \hpr^{-1}\:.
\ee
We can check that the corresponding  Hamiltonian operator:
\be
\hh=\int d^{3}x\: \qty[ \partial^{0}\hphd (x)\partial_{0}\hph (x)-\partial^{i}\hphd (x)\partial_{i}\hph (x)+\hphd (x)M^{2}\hph (x)]
\ee
is also $\hpr$-pseudo-Hermitian, i.e., $\hh^{\dag}=\hpr \hh\hpr^{-1}$.

However, for a non-Hermitian Hamiltonian, the standard scalar/pseudoscalar parity transformation is incorrect. This can be seen implicitly by parity-transforming Hamilton's equation:
\be
\begin{aligned}
    P: [\hph (x),\hh]=i\der{0}{}{\hph}(x) \mapsto [\hpr\hph (x)\hpr^{-1},\hpr\hh \hpr^{-1}]=i\hpr\der{0}{}{\hph}(x)\hpr^{-1}\\ \implies [P\hph(x^{P}),\hh^{\dag}]=iP\der{0}{}{\hph}(x^{P})\;.
\end{aligned}
\ee
Multiplying the left-hand side by $P^{-1}$, and since the Hamiltonian $\hh$ is independent of $x$, we can relabel $x^{P}$ to $x$, giving
\be
P: [\hph (x),\hh]=i\der{0}{}{\hph}(x) \longmapsto [\hph (x),\hh^{\dag}]=i\der{0}{}{\hph}(x)\;.
\ee
By virtue of the non-Hermiticity of the Hamiltonian, $\hh^{\dag}\neq \hh$, we see that the field operator cannot transform under parity in the usual way. 

We can also see this explicitly by acting with parity on the momentum decomposition of the field $\hph$. Varying the Lagrangian with respect to $\hphd$, gives the Euler-Lagrange equation for $\hph$:
\be\label{eq:phi_eq_of_motion}
\partial^{\mu}\partial_{\mu}\hph (x)+M^{2}\hph (x)=0\;.
\ee
The general solution to this equation can be written in momentum space as
\be\label{eq:momentum_decomposition_phi}
\hph (x)=\int \frac{d^3\vp}{ (2\pi)^{3}}  (2E_{\vp})^{-\hf} [e^{-ip\cdot x}\hat{a} (0,\vp)+e^{ip\cdot x}\hat{c}^{\dag} (0,\vp)]\;,
\ee
where $E_{\vp}=\sqrt{\vp^{2}\id_{2}+M^{2}}$ is the $2\times2$ energy-momentum matrix, which is nondiagonal and non-Hermitian due to the non-Hermiticity of $M^{2}$. Here, $\hat{a}$ and $\hat{c}^{\dag}$ are the $2$-component annihilation and creation operators. 

As the non-Hermiticity comes purely from the mass matrix, it affects the $\mu=0$ component of the 4-momentum, i.e., the energy $p^{0}=E_{\vp}$, but not the 3-momentum $\vp$. At $t=0$, the annihilation $\hat{a} (0,\vp)$ and creation $\hat{c}^{\dag} (0,\vp)$ operators do not depend on the energy and only on the 3-momentum. Thus, at $t=0$, they are not affected by the non-Hermiticity of the Lagrangian and should transform under parity and time reversal in the standard way:
\begin{gather}
\begin{array}{ll}\label{eq:parity_on_ca_operators}
    \hpr \ha{i}{\vp}\hpr^{-1}=P^{ij}\:\ha{j}{-\vp}\:, & \:\:
    \hpr \hdc{i}{\vp}\hpr^{-1}=P^{ij}\:\hdc{j}{-\vp}\:, \\
    \hpr \hda{i}{\vp}\hpr^{-1}=\hda{j}{-\vp}\:P^{ji}\:, &\:\:
    \hpr \hc{i}{\vp}\hpr^{-1}=\hc{j}{-\vp}\:P^{ji} \:.
    \end{array}\\ 
\begin{array}{ll}
    \htr \ha{i}{\vp}\htr^{-1}=\ha{i}{-\vp}\:, &\:\:
    \htr \hdc{i}{\vp}\htr^{-1}=\hdc{i}{-\vp}\:, \\
    \htr \hda{i}{\vp}\htr^{-1}=\hda{i}{-\vp}\:, &\:\:
    \htr \hc{i}{\vp}\htr^{-1}=\hc{i}{-\vp}\:.
    \end{array}
\end{gather}
Here, $\hat{a}^{1}$ and $\hat{c}^{\dag 1}$ transform as scalars with positive intrinsic parity $+1$, whereas $\hat{a}^{2}$ and $\hat{c}^{\dag 2}$ transform as pseudoscalars with negative intrinsic parity $-1$. Moreover, at $t=0$, they should have the usual commutation relations:
\be
\begin{aligned}
    &[\ha{i}{\vp},\hda{j}{\vec{q}}]=
    [\hc{i}{\vp},\hdc{j}{\vec{q}}]=\delta\indices{^i^j}(2\pi)^{2}\delta^{(3)}(\vp-\vec{q})\;,\\
    &[\ha{i}{\vp},\hc{j}{\vec{q}}]=[\ha{i}{\vp},\hdc{j}{\vec{q}}]=0\;.
\end{aligned}
\ee
Hence, if we act with parity on the momentum decomposition of the field $\hph$ in Eq.~\eqref{eq:momentum_decomposition_phi}, we find that its parity transformation is indeed nontrivial:
\be\label{eq:parity_transformation_of_phi}
\hpr\hph (x)\hpr^{-1}=P\int \frac{d^3\vp}{ (2\pi)^{3}}  (2E^{\dag}_{\vp})^{-\hf} [e^{-ip^\dag \cdot x^{P}}\hat{a} (0,\vp)+e^{ip^\dag \cdot x^{P}}\hat{c}^{\dag} (0,\vp)]\neq P\:\hph (x^{P})\;.
\ee

Now consider the Euler-Lagrange equations of motion. If we vary the Lagrangian with $\hph$, then we get the Euler-Lagrange equation for $\hphd$:
\be
\partial^{\mu}\partial_{\mu}\hphd (x)+\hphd (x)M^{2}=0\;.
\ee
However, it is not the Hermitian conjugate of the Euler-Lagrange equation for $\hph$ given in Eq.~\eqref{eq:phi_eq_of_motion} by virtue of the non-Hermiticity of the mass mixing matrix, $M^{2^\dag}\neq M^{2}$:
\be
(\partial^{\mu}\partial_{\mu}\hphd (x)+\hphd (x)M^{2})^{\dag}=0 \implies 
\partial^{\mu}\partial_{\mu}\hph (x)+M^{2^\dag}\hph (x)=0\:.
\ee

These issues are a direct consequence of the fields $\hph$ and $\hphd$ being governed by two different Hamiltonians, $\hh$ and $\hh^{\dag}$, respectively. As described in the preceding section, we fix this by determining the dual field $\hat{\td{\phi}}^{\dag}$ and using it to construct a consistent Lagrangian operator.

\subsection{Poincar\'{e}-invariant Lagrangian}\label{sec:correct_lagrangian}

Our prescription for a self-consistent non-Hermitian quantum field theory suggests that we should begin with a Lagrangian built from the field operator $\hph$ and its dual $\hat{\td{\phi}}^{\dag}$:
\be\label{eq:correct_quantum_Lagrangian}
\hlag (x)=\der{\mu}{}{\thph{\dag}{}{x}}\der{\mu}{}{\hph (x)}-\thph{\dag}{}{x} M^{2}\hph (x)\:.
\ee
Here, we have the same non-Hermitian mass mixing matrix as in Eq.~\eqref{eq:mass_mixing_matrix}.
The dual field for an $n$-component scalar field is given by Eq.~\eqref{eq:dual_field_for_n_comp_scalars}. In this example, $n=2$, and the dual field is
\be
\thph{\dag}{}{x}=\heta^{-1} \hphd (x^{\eta})\heta \:\Pi \:,\text{\:\:\: with\:\:\:} \Pi=\mqty(\pi_{1}&0\\0&\pi_{2})\:.
\ee
At this point, the operator $\heta$ and the matrix $\Pi$ are yet to be determined. 

By construction, this Lagrangian is $\heta$-pseudo-Hermitian and the mass matrix is $\Pi$-pseudo-Hermitian:
\be
\hlag^{\dag}(x^{\eta})=\heta \hlag (x)\heta^{-1}\text{\:\:\:and\:\:\:}M^{2^\dag}=\Pi M^{2}\Pi^{-1}\:.
\ee
In the previous subsection, we found that the \comma\comma naive'' Lagrangian and the Hamiltonian operator are $\hpr$-pseudo-Hermitian with respect to the parity operator $\hpr$ and the mass matrix is $P$-pseudo-Hermitian with respect to the parity matrix $P$ given in Eq.~\eqref{eq:parity_matrix}. Hence, this suggests natural choices for the operator $\heta$ to be the parity operator $\hpr$ and $\Pi$ to be the parity matrix $P$. 

Indeed, if we vary the Lagrangian in Eq.~\eqref{eq:correct_quantum_Lagrangian} with $\hph$, we get the Euler-Lagrange equation for the dual field $\hat{\td{\phi}}^{\dag}$:
\be
\partial^{\mu}\partial_{\mu}\thph{\dag}{}{x}+\thph{\dag}{}{x}M^{2}=0\:.
\ee
In momentum space, it has a solution
\be
\thph{\dag}{}{x}=\int \frac{d^3\vp}{ (2\pi)^{3}} [\hat{a}^{\dag} (0,\vp) e^{i p\cdot x}+\hat{c} (0,\vp) e^{-i p\cdot x}] (2E_{\vp})^{-\hf}\:.
\ee
This is exactly the parity transformation of $\hph$ found in Eq.~\eqref{eq:parity_transformation_of_phi}, which can be rewritten as
\be
\thph{\dag}{}{x}=\hpr^{-1}\hphd (x^{P})\hpr \: P\:.
\ee
We note that this matches the construction proposed in Ref.~\cite{Alexandre:2020gah}. Given the dual field above, the Lagrangian in Eq.~\eqref{eq:correct_quantum_Lagrangian} operator is indeed $\hpr$-pseudo-Hermitian:
\be
\hlag^{\dag} (x^{P})=\hpr\hlag (x)\hpr^{-1}\:.
\ee

Having chosen $\heta$ to be the parity operator $\hpr$, a natural choice for the inner product yielding real eigenspectra would be $\braket{\cdot}{\hpr \cdot}$. However, if we wish for the Fock-space 3-momentum operator to remain Hermitian $\hat{\vec{P}}^{\dag}=\hat{\vec{P}}$, we must recognize that the inner product $\braket{\cdot}{\hpr \cdot}$ will not be invariant under space translations:
\be
\bra{\al}\hpr\ket{\bt}\mapsto \bra{\al}e^{-i\ep_{a}\hat{P}^{\dag^a}}\hpr e^{i\ep_{a}\hat{P}^{a}}\ket{\bt}=\bra{\al}\hpr e^{2i\ep_{a}\hat{P}^{a}}\ket{\bt}\;.
\ee
This is because the parity operator flips the sign of the 3-momentum, i.e.,
\be
\hpr \hat{\vec{P}}\hpr\ket{\vp,k}=\hpr \hat{\vec{P}} P^{jk}\ket{-\vp,j}= (-\vp)P^{jk}\hpr\ket{-\vp,j}=(-\vp)P^{jk}P^{ij}\ket{\vp,i}=-\vp\ket{\vp,k}\;,
\ee
so that
\be
\hpr \hat{\vec{P}}\hpr=-\hat{\vec{P}}\:.
\ee
As we established in Eq.~\eqref{eq:non_Hermiticity_of_other_generators}, to preserve Poincar\'e invariance, all of the generators must be $\heta$-pseudo-Hermitian. Hence, they are Hermitian if and only if they commute with $\heta$. However, if we wish to keep the $3$-momentum operator Hermitian, it will not commute with the parity operator, and the theory with $\heta=\hpr$ will then not be Poincar\'e invariant.

\subsection{C operator}\label{sec:C_Operator}

In $PT$-symmetric quantum mechanics, where the Hamiltonian is non-Hermitian, but $PT$-symmetric, i.e., $[H,PT]=0$, we discover an additional discrete symmetry, denoted as $C$ such that $[H,C]=0$ and $C^2=\id$~\cite{Bender:2002vv}. Despite its name, this symmetry is unrelated to charge conjugation. In the $PT$-unbroken phase, where the Hamiltonian has real eigenvalues, we use this symmetry to construct a positive-definite metric $PC$. This metric ensures that the eigenstates of the Hamiltonian are orthogonal with respect to the positive-definite inner product $\braket{\cdot}{PC\cdot}$.

It turns out that such a symmetry exists in any discrete pseudo-Hermitian system, provided it is diagonalizable. In our example, this system is the $P$-pseudo-Hermitian mass mixing matrix. It is diagonalizable with an eigenspectrum $M^{2}\ket{\psi_{1,2}}=\bar{m}^{2}_{1,2}\ket{\psi_{1,2}}$, where $\bar{m}_{1}$ is the \comma\comma$-$'' and $\bar{m}_{2}$ is the \comma\comma$+$'' root, respectively~\cite{Alexandre:2017foi}:
\be
\bar{m}^{2}_{1,2}=\frac{m^{2}_{2}+m^{2}_{1}}{2}\mp\frac{m^{2}_{1}-m^{2}_{2}}{2}\sqrt{1-\nu^{2}}\text{\:\:\:with\:\:\:}\nu\df\frac{2\mu^{2}}{m_{1}^{2}-m^{2}_{2}}\:.
\ee
We have chosen $m_{1}^{2}>m^{2}_{2}$ so that $\nu>0$. The mass eigenvalues $\bar{m}_{1,2}\in\rn$ are real for $\nu\leq 1$ and the mass eigenstates will be \comma\comma pseudo'' orthonormal with respect to the indefinite inner product $\braket{\cdot}{\cdot}_{P}=\braket{\cdot}{P\cdot}$:
\be
\braket{\psi_{m}}{\psi_{n}}_{P}=(-1)^{m}\delta_{mn}\text{\:\:\:for\:\:\:}m,n\in\{1,2\}\:.
\ee
Here, the normalized mass eigenstates are~\cite{Alexandre:2017foi}
\be
\ket{\psi_{1}}=N\mqty(-1+\sqrt{1-\nu^{2}}\\ \nu)\:\text{\:\:\:and\:\:\:}
\ket{\psi_{2}}=N\mqty(\nu\\-1+\sqrt{1-\nu^{2}})\:,
\ee
with normalization constant $N=\qty[2(\sqrt{1-\nu^{2}}-(1-\nu^{2}))]^{-\hf}$.

As mentioned, this system will have an additional discrete symmetry $[M^{2},C]=0$ with $C^{2}=\id_{2}$. While this symmetry exists for both real and and complex mass eigenvalues, in the case of real eigenspectrum, it will be related to the parity matrix via~\cite{Alexandre:2020gah}
\be\label{eq:operator_C}
C=\displaystyle\sum_{n=1}^{2}\ket{\psi_{n}}\otimes\bra{\psi_{n}}\: P=\frac{1}{\sqrt{1-\nu^{2}}}\mqty(1&\nu\\-\nu&-1)\text{\:\:\:and\:\:\:} PC=\frac{1}{\sqrt{1-\nu^{2}}}\mqty(1&\nu\\ \nu & 1)\:.
\ee
\noindent Hence, the mass eigenstates will be orthogonal with respect to a positive-definite inner product $\braket{\cdot}{\cdot}_{PC}=\braket{\cdot}{PC\cdot}$:
\be
\braket{\psi_{m}}{\psi_{n}}_{PC}=\delta_{mn}\text{\:\:\:for\:\:\:}m,n\in\{1,2\}\:,
\ee

As the pseudo-Hermiticity of our Lagrangian arises entirely from the $P$-pseudo-Hermitian mass mixing matrix,
the $\hpr$-pseudo-Hermitian Hamiltonian operator $\hh$ is also diagonalizable. Thus, it also contains a discrete symmetry  $[\hh,\hC]=0$ with  $\hC^{2}=\hat{\id}$. A difficult, but possible, way to find this symmetry is to take an infinite sum of multiparticle eigenstates of $\hh$ similar to Eq.~\eqref{eq:operator_C}. However, a more straightforward approach would be to notice that the $C$ matrix is just the parity matrix $P$ in the basis where $M^{2}$ is nondiagonal~\cite{Alexandre:2018uol}:
\be
M^{2}=R D^{2} R^{-1}\text{\:\:\:with\:\:\:}R=N\mqty(-1+\sqrt{1-\nu^{2}}&\nu\\ \nu & -1+\sqrt{1-\nu^{2}})\:.
\ee
In the eigenbasis of $M^{2}$, $C$ is just the parity matrix $P$~\cite{Alexandre:2018uol}:
\be
C=RPR^{-1}\:.
\ee
Hence, the $\hC$ operator is just the parity operator $\hpr$ in the eigenbasis of the Hamiltonian operator $\hh$:
\be
\hh=\hat{R}\hat{D}\hat{R}^{-1}\longrightarrow \hC=\hat{R}\hpr\hat{R}^{-1}\:,
\ee
where $\hat{D}:\fs\mapsto\fs$ is the diagonalized Hamiltonian operator.

Since, the diagonalized mass mixing matrix commutes with parity, i.e., $[D,P]=0$, so does the diagonalized Hamiltonian operator $[\hat{D},\hpr]=0$. It follows that $\hC$ acts on the annihilation/ creation operators just as the parity operator $\hpr$, but with $C$ instead of $P$ as given in Eq.~\eqref{eq:parity_on_ca_operators}:
\be\label{eq:C_on_creation_annihilation_operators}
\begin{aligned}
\hC\ha{}{\vp}\hC^{-1}=C\ha{}{-\vp}\:, \:\:\:\:\:\:\: \hC\hdc{}{\vp}\hC^{-1}=C\hdc{}{-\vp}\:,\\ 
\hC\hda{}{\vp}\hC^{-1}=\hda{}{-\vp} C\:,
\:\:\:\:\:\: \hC\hc{}{\vp}\hC^{-1}=\hc{}{-\vp} C\;.
\end{aligned}
\ee
Indeed, if we expand the Hamiltonian in terms of creation and annihilation operators, then we find that it commutes with $\hC$:
\be
\hh=\int\frac{d^{3}p}{(2\pi)^{3}}\:\qty[\hda{i}{\vp}E^{ij}_{\vp} \ha{j}{\vp}+\hdc{i}{\vp}E^{ij}_{\vp} \hc{j}{\vp}] \implies \hC\hh\hC^{-1}=\hh\;.
\ee
This is due to the matrix $C$ commuting with $E_{\vp}=\sqrt{\vec{p}^{2}\id_{2}+M^{2}}$. Also, the operator $\hC$ will square to unity: 
\be
\hC^{2}\ha{}{\vp}\hC^{-2}=\ha{}{\vp}\implies \hC^{2}=\hat{\id}\;.
\ee

The easiest way to find the operators $\hpr$ and $\hC$ is to use the Baker-Campbell-Hausdorf formula:
\begin{align}
\hpr&=\text{exp}\qty[-i\frac{\pi}{2} \int \frac{d^{3}p}{(2\pi)^{3}} \qty(\hat{a}^{\dag i}_{\vp}\: P^{ij}\: \hat{a}^{j}_{\vp}-\hat{a}^{\dag i}_{\vp}\hat{a}^{i}_{-\vp} +\hat{c}^{i}_{\vp} \:P^{ij}\:\hat{c}^{\dag j}_{\vp}-\hat{c}^{i}_{\vp}\hat{c}^{\dag i}_{-\vp} ) ]\:,\\
\hC&=\text{exp}\qty[-i\frac{\pi}{2} \int \frac{d^{3}p}{(2\pi)^{3}} \qty(   
\hat{a}^{\dag i}_{\vp}\: C^{ij}\: \hat{a}^{j}_{\vp}-\hat{a}^{\dag i}_{\vp}\hat{a}^{i}_{-\vp} +\hat{c}^{i}_{\vp} \:C^{ij}\:\hat{c}^{\dag j}_{\vp}-\hat{c}^{i}_{\vp}\hat{c}^{\dag i}_{-\vp} ) ]\:,
\end{align}
where $\hat{a}_{\vp}\df\ha{}{\vp}$ and $\hat{c}_{\vp}\df\hc{}{\vp}$ are creation/annihilation operators at $t=0$.

The important property of the $\hC$ operator is that it flips the sign of the 3-momentum operator:
\be
\hC\hat{\vec{P}}\hC \ket{\vp,k}=\hC \hat{\vec{P}} C^{jk}\ket{-\vp,j}= (-\vp)C^{jk}\hC\ket{-\vp,j}=(-\vp)C^{jk}C^{ij}\ket{\vp,i}=-\vp\ket{\vp,k}\:,
\ee
so that
\be
\hC \hat{\vec{P}}\hC=-\hat{\vec{P}}\;.
\ee
As noted in the previous subsection, the parity operator also flips the sign of the 3-momentum operator. Thus, if we wish to keep the 3-momentum operator Hermitian, the theory with $\heta=\hpr$ will not be Poincar\'e invariant. However, the existence of the discrete symmetry $\hC$, allows us to construct a new positive-definite metric operator $\hpr\hC$, such that the eigenstates of the Hamiltonian operator are orthogonal with respect to the positive-definite inner product $\braket{\cdot}{\cdot}_{\hpr\hC}=\braket{\cdot}{\hpr\hC\cdot}$ and have positive norms. But more importantly, if we choose $\heta=\hpr\hC$ it commutes with the 3-momentum operator:
\be
 (\hpr\hC)\hat{P}^{i} (\hpr\hC)^{-1}=\hat{P}^{i}\;.
\ee
Hence, the 3-momentum can simultaneously be Hermitian and $\hpr\hC$-pseudo-Hermitian, leaving the theory Poincar\'e invariant.

Therefore, in Eq.~\eqref{eq:dual_field_for_n_comp_scalars}, we take $\heta=\hpr\hC,\:\Pi=PC$ and the coordinates remain unchanged under this transformation, i.e., $x^{PC}=x$, so that the dual field is
\be
\thph{\dag}{}{x}=(\hpr\hC)^{-1}\hphd (x)(\hpr\hC)PC\:.
\ee
Both the Lagrangian and the Hamiltonian are $\hpr\hC$-pseudo-Hermitian:
\be
\hlag^{\dag}(x)=(\hpr\hC) \hlag(x) (\hpr\hC)^{-1}\text{\:\:\:and\:\:\:}\hh^{\dag}=(\hpr\hC)\hh(\hpr\hC)^{-1}\:.
\ee
The inner product $\braket{\cdot}{\hpr\hC\cdot}$ yields a theory that is invariant under proper Poincar\'e transformations.


\section{Conclusion}
\label{sec:conclusions}

In this work, we have considered a generalization of the Poincar\'{e} group when the generator of time translations, i.e., the Hamiltonian operator, is non-Hermitian. The time evolution of the quantum field operator $\hps$ and its Hermitian conjugate $\hpsd$ are then governed by distinct Hamiltonians, $\hh$ and $\hh^{\dag}$, respectively. As a consequence, a theory built from $\hps$ and $\hpsd$ exhibits inconsistent equations of motion and lacks Poincar\'e invariance. We have also shown that when the Hamiltonian is non-Hermitian, its non-Hermiticity extends to the other group generators, such as space translations, rotations, and boosts. Specifically, if the Hamiltonian is diagonalizable, we can always find an operator $\heta$ such that $\hh^{\dag}=\heta \hh \heta^{-1}$, which implies that the Hamiltonian is $\heta$-pseudo-Hermitian. The pseudo-Hermiticity of the Hamiltonian leads to pseudo-Hermiticity of the other generators in the Poincar\'e algebra. These generators become Hermitian if and only if they commute with $\heta$.

For a quantum field theory to be Poincar\'e invariant, the Lagrangian should transform as a single object under the proper Poincar\'e group. However, we observe that the field operator $\hps$ and its Hermitian conjugate $\hpsd$ transform in two different representations, leading to the lack of Poincar\'e invariance in both the Lagrangian and the theory as a whole. This prompts us to search for a new conjugate field, which we refer to as the \comma`dual'' field. This dual field operator, denoted by $\hat{\td{\psi}}^{\dag}$, transforms in the dual representation of $\hps$. In Sec.~\ref{sec:Fock_space_reps}, we found it to have a general form, which holds for fields of any spin $j$:
\be
\label{eq:mainresult}
\hat{\td{\psi}}^{\dag}_{j}(x)=\heta^{-1}\hat{\psi}^{\dag}_{j}(x^{\eta})\heta \:\pi\;.
\ee
Herein, the operator $\heta$ is such that the Hamiltonian is $\heta$-pseudo-Hermitian, $x^{\eta}$ is the coordinate transformation with respect to the operator $\heta$ (e.g., parity $x^{P}$), and the matrix $\pi$ is such that the finite-dimensional Lorentz representations are $\pi$-pseudo-Hermitian. Equation~\eqref{eq:mainresult} is the central result of this work, and we have demonstrated its importance using a simple model of two complex scalar fields with non-Hermitian mass mixing, where the Hamiltonian is pseudo-Hermitian with respect to the parity operator $\hpr$. 

With this result, we have established a fundamental framework for developing self-consistent non-Hermitian quantum field theories. By laying down these foundational principles, we pave the way for future applications, in particular in the challenging context of interacting non-Hermitian quantum field theories.

\vspace{12pt}
\noindent\emph{Note added}. After the preprint of this work was posted, Ref.~\cite{Ferro-Hernandez:2023ymz} appeared, which brought the earlier work~\cite{LeClair:2007iy} to our attention. The construction of the dual field for the second-order fermionic theory described in these works, while not motivated by Poincar\'{e} invariance, bears some similarities with the construction described here, and we leave a detailed application of the present approach to this theory for future work.


\begin{acknowledgments}

The authors thank Maxim Chernodub, Jeff Forshaw and Viola Gattus for helpful discussions. This work was supported by the University of Manchester; the Science and Technology Facilities Council (STFC) [Grant No.~ST/X00077X/1]; and a United Kingdom Research and Innovation (UKRI) Future Leaders Fellowship [Grant No.~MR/V021974/1 and No.~MR/V021974/2]. The authors also acknowledge support during the early stages of this work from the University of Nottingham and a Nottingham Research Fellowship.

\end{acknowledgments}

\section*{Data Access Statement}

No data were created or analysed in this study.



\end{document}